%% file: finaleta.tex
\documentclass[11pt,a4paper]{article}
\usepackage{lmodern}

\usepackage[T1]{fontenc}
\usepackage[utf8]{inputenc}
\usepackage{color}
\usepackage{amsmath}
\usepackage{amssymb}
\usepackage{graphicx}
\usepackage{esint}
\usepackage{float}

\usepackage{jcappub}

\title{\boldmath Model-independent reconstruction of the linear anisotropic stress $ \eta$}

\author[a]{Ana Marta Pinho}
\author[a,b]{Santiago Casas}
\author[a]{Luca Amendola}

\affiliation[a]{ITP, Ruprecht-Karls-Universität Heidelberg \\ Philosophenweg 16, 69120 Heidelberg, Germany}
\affiliation[b]{AIM, CEA, CNRS, Université Paris-Saclay, Université Paris Diderot, \\ 
Sorbonne Paris Cité, F-91191 Gif-sur-Yvette, France
}

\emailAdd{pinho@thphys.uni-heidelberg.de}
\emailAdd{santiago.casas@cea.fr}
\emailAdd{amendola@thphys.uni-heidelberg.de}

\abstract{
In this work, we use recent data on the Hubble expansion rate $H(z)$, 
the quantity $f\sigma_8(z)$ from redshift space distortions and the statistic 
$E_g$ from clustering and lensing observables to constrain
in a model-independent way  the linear anisotropic stress parameter
$\eta$. This estimate is free of assumptions about initial conditions, bias,
 the abundance of dark matter and the background expansion. We denote this observable estimator
as $\eta_{{\rm obs}}$. If $\eta_{{\rm obs}}$ turns out to be different
from unity, it would imply either a modification of gravity or a non-perfect
fluid form of dark energy clustering at sub-horizon scales. Using
three different methods to reconstruct the underlying model from data, 
we report the value of $\eta_{{\rm obs}}$ at  three redshift values, $z=0.29, 0.58, 0.86$.
Using the method of polynomial regression, we find $\eta_{{\rm obs}}=0.57\pm1.05$, 
$\eta_{{\rm obs}}=0.48\pm0.96$, and $\eta_{{\rm obs}}=-0.11\pm3.21$, respectively. 
Assuming a constant $\eta_{{\rm obs}}$ in this range, we find  $\eta_{{\rm obs}}=0.49\pm0.69$. 
We consider this method as our fiducial result, for reasons clarified in the text.
The other two methods give for a constant anisotropic stress $\eta_{{\rm obs}}=0.15\pm0.27$
(binning) and $\eta_{{\rm obs}}=0.53 \pm 0.19$ (Gaussian Process). 
We find that all three estimates are compatible with each other within
their $1\sigma$ error bars. While the polynomial regression method is 
compatible with standard gravity, the other two methods are in tension with it. 
}

\keywords{Model-independent approach - Cosmology - Gravity}

\arxivnumber{arXiv:1805.00027}

\begin{document}

\maketitle
 
\section{Introduction}

The recent observation of gravitational waves from a neutron star
merger GW170817 by the LIGO/VIRGO collaboration together with its
electromagnetic counterpart GRB170817 observed immediately afterwards
by several telescopes around the world \cite{Abbott2017,Abbott2017search,Albert2017,Abbott2017dynamical,Abbott2017multi},
placed very tight constraints on the difference between the speed
of gravitational waves $c_{T}$ and the speed of light $c$, constraining
it to be fractionally smaller than $10^{-15}$ \cite{Abbott2017}. Consequences of such measurement (as for instance discussed in \cite{Lombriser2015, Lombriser2016}) include ruling out a sector of Horndeski's theory \cite{Ezquiaga2017,Sakstein2017,Creminelli2017,Baker2017,Amendola:2017orw}, that is
the most general theory of a single scalar field having only second
order equations of motion and being free of ghost instabilities \cite{Horndeski1974}.
However, to rule out these sectors of the Horndeski Lagrangian one
has to assume no extreme fine-tuning, the absence of attractors (see e.g. \cite{Amendola2013}), and
a universal coupling between matter and the scalar field. Even after the $c_T$ constraint, therefore, the Horndeski
Lagrangian remains  the most interesting extension of Einstein's
gravity to test in cosmology.

In the general Horndeski theory, and also in bimetric gravity \cite{Konnig2014},
one can show that the gravitational slip, defined as the ratio of
the gravitational potentials $\eta=-\Phi/\Psi$, has a relatively
simple functional form in Fourier space 
\begin{equation} \label{eq:eta-quasistat}
\eta=h_{2}\frac{1+h_{4}k^{2}}{1+h_{5}k^{2}}\,\,,
\end{equation}
with coefficients $h_{i}$ which are free functions of time and depend
on the four free functions appearing in the Horndeski Lagrangian
\cite{Amendola2012,DeFelice2011,Felice2012}. The constraint from gravitational waves sets $h_{2}$ equal
to 1 since $h_2 = 1/c_T^{2}$ \cite{Amendola2012}, but leaves free all the other functions. In the limit of large (but still sub-horizon)
scales and provided that the theory does contain at least one mass scale
besides the Planck mass \cite{Nersisyan:2018auj}, one obtains $\eta = 1$. In all other cases, $\eta\ne 1$ signals  a  deviation from standard gravity or a form of dark matter that cannot be approximated by a perfect fluid.

Considering model-independent observables and linear structure formation, and assuming gravity
remains universally coupled also when modified,
one can build an estimate of $\eta$ formed by three directly observable
functions of redshift which we denote $E(z)$, $P_{2}(z)$ and $P_{3}(z)$
(see \cite{Motta2013}). They will be defined
in detail in the following section. The first function, $E(z)$, is
the dimensionless Hubble function. The second one, $P_{2}(z)$, is
equivalent  to the $E_{G}(z)$ statistics \cite{Leonard2015}, that depends on
the lensing potential and on  the growth rate of structure formation.
Finally, $P_{3}(z)$ (introduced in \cite{Amendola2012}) is related to the derivative of $f\sigma_{8}(z)$, which is the growth rate of
matter density perturbations times the normalization of the power spectrum. 
This is measured
by galaxy clustering using redshift space distortions.

In order to reconstruct $E(z)$, $P_{2}(z)$ and $P_{3}(z)$, we use
the most recent data available for $H(z)$ obtained with Type Ia Supernovae
and cosmic chronometers, while for $f\sigma_{8}(z)$ and $E_{G}$ we
employ redshift space distortion and galaxy-galaxy lensing data from several collaborations, listed below in section \ref{datapart}.

The problem of reconstructing an unknown function and its
derivative from sparse and noisy data is not trivial and it is an
important task in all areas of science. In this work, we use three
different strategies to estimate the unknown functions and their derivatives from the data.
As the first method, we use a \textit{\emph{simple}} binning
formalism, in which we group the available data in redshift bins and
use discrete finite differences to compute the derivatives at the
corresponding redshifts. This method suffers from strong numerical
uncertainties since the derivatives are very sensitive to the binning
size and the method cannot capture high-frequency modes in the data. The second one is the Gaussian Process
method, a generalization of a Gaussian distribution, where instead
of random variables, one has a distribution of random functions, connected
by a specific correlation function. This method has been used several
times in cosmology, especially for the determination of the equation
of state of dark energy $w$ and the Hubble function $H(z)$ (see
\citep{Seikel2012,Yu2017,Gomez-Valent2018,Melia2018,Holsclaw2011}). The third method consists of a polynomial 
regression (used for example recently in \cite{Marra2017}), in which
one assumes a linear model for the underlying function. Using the so-called
\textit{normal equation}, we reconstruct the coefficients
of the polynomial, which represents our continuous interpolation function of the data, which is later evaluated at specific redshifts.

In section \ref{theopart} we explain the theoretical foundations
of the determination of $\eta$. Section \ref{datapart} describes
the data used in our analysis and their processing before we apply
our three reconstruction methods which are explained in section \ref{recopart}.
The estimation of the gravitational slip and overall discussion of
our results can be found in section \ref{resupart}.
Finally, we present some of the caveats of the methods and suggest
ways  to improve this analysis with future work.

\section{Model-independent observables}

\label{theopart}

The geometry of the Universe can be well described by small scalar
perturbations around a flat FLRW metric $ds^{2}=-(1+2\Psi)\textrm{d}t^{2}+a(t)^{2}(1+2\Phi)\textrm{d}\mathbf{x}^{2}$,
with scale factor $a$ and two scalar gravitational potentials $\Psi$
and $\Phi$. Using Einstein's field equations and a presureless perfect
fluid for matter, we can derive the two Poisson equations in Fourier
space 
\begin{align}
-k^{2}(\Psi-\Phi) & =4\pi GH(z)^{-2}\Sigma(k,z)\rho_{m}(z)\delta_{m}(z,k)\,\label{eq:poisson}\\
-k^{2}\Psi & =4\pi GH(z)^{-2}\mu(k,z)\rho_{m}(z)\delta_{m}(z,k)\,\label{eq:poisson2}
\end{align}
where $z$ is the redshift, $k$ the scale in terms of the cosmological
horizon (the comoving scale $k_{com}$ divided by $aH$), $\rho_{m}$
is the background average matter density of the Universe, $\delta_{m}$
the matter density contrast and $\Sigma$ and $\mu$ are two functions
of scale and time which quantify the departure from standard gravity.
In Einstein's General Relativity, these functions reduce to to $\Sigma=2$
and $\mu=1$. The gravitational slip $\eta$ is defined as the ratio
between the two gravitational potentials 
\begin{equation}
\eta=-\Phi/\Psi\,\,,
\end{equation}
where the perturbation variables are considered to be the root-mean-squares of the corresponding random variables.
Taking the appropriate ratios of the Poisson equations (Eq.~\ref{eq:poisson} and Eq.~\ref{eq:poisson2}) defined above, 
we find a simple relation for the modified lensing parameter: $\Sigma=\mu(1+\eta)$.
If we make no assumptions about the initial conditions of the Universe,
neither on the primordial power spectrum, nor on the nature
of dark matter or the details of galaxy bias, we cannot determine
the matter background density nor the matter overdensity in a model-independent
way (see \cite{Amendola2012}). Therefore, a quantity like $\mu(k,z)$
in modified gravity cannot be estimated without first assuming a model. However, one can define model-independent observable
quantities which do not depend on the aforementioned assumptions.
Following \cite{Motta2013}, these variables are called $A$, $R$,
$L$ and $E$, respectively denoting amplitude, redshift-space distortions, lensing,
and the dimensionless Hubble function. They are defined as 
\begin{equation}
\begin{aligned}
A & = b \delta_{m}\,, \qquad & R & = f\delta_{m},\\
L & =\Omega_{m0}\Sigma\delta_{m} \,,\qquad & E & = H/H_{0} \;.
\end{aligned}
\label{eq:arle}
\end{equation}
where $b$ is the galaxy-matter linear bias, $f=\delta_m'/\delta_m$ is the growth rate where the prime is derivative with respect to $\ln a$, and $\Omega_{m0}$ is today's matter fractional density. Both $f$ and $b$ can be in general time- and space-dependent. The formalism below can be applied also in this case, but since the available data do not provide the space dependence, in the following we will assume that it can be ignored. For the same reason, also $\eta$ will  be assumed to be independent of scale in the observed range. Looking at Eq. (\ref{eq:eta-quasistat}), one sees that scale-independence sets in  either at small scales $k\gg 1$, or at large scales $k\ll 1$ (but in  this case $\eta\to 1$) or at all scales if $h_4=h_5$ or if the theory does not contain a mass scale.

With the definitions (\ref{eq:arle}), it was shown in
\cite{Amendola2012,Motta2013} that one can obtain three quantities
which are model-independent and cancel out the effects of the shape
of the primordial power spectrum and the galaxy bias, namely 
\begin{align}
P_{1} & \equiv\frac{R}{A}=\frac{f}{b},\\
P_{2} & \equiv\frac{L}{R}=\frac{\Omega_{\text{m}0}\Sigma}{f},\\
P_{3} & \equiv\frac{R'}{R}=f+\frac{f'}{f}=\frac{(f\sigma_{8}(z))'}{f\sigma_{8}(z)}\,\,\,.
\end{align}
We have defined
$f\sigma_{8}(z)$ as 
\begin{equation}
f\sigma_{8}(z)=\sigma_{8}G(z)f(z)\,,
\end{equation}
where $\sigma_{8}$ is the amplitude of the linear power spectrum
defined in a spherical shell of radius $8\textrm{ Mpc}$ at redshift
$z=0$ and $G(z)$ is the growth function normalized to unity today, $\delta(z)=\delta_{m,0}G(z)$. 

Together with the continuity equation and the Euler equation, relating
the divergence $\theta$ of the peculiar velocities of galaxies  propagating on geodesics
to the gravitational potential $\Psi$ 
\begin{equation}
(a^{2}\theta)'=a^{2}k^{2}H\Psi
\end{equation}
we can write down the lensing and Poisson equations in Fourier space,
respectively, in the following way 
\begin{align}
-k^{2}(\Psi-\Phi) & =\frac{3(1+z)^{3}L}{2E^{2}}\,\label{eq:lensing-arle}\\
-k^{2}\Psi & =R'+R\bigg(2+\frac{E'}{E}\bigg)\,.\label{eq:fgrowth-arle}
\end{align}
This last equation is usually known as the equation of linear growth
of matter perturbations. 
Dividing the lensing equation Eq.~(\ref{eq:lensing-arle}) 
by the equation for the growth of
structure Eq.~(\ref{eq:fgrowth-arle}), we can obtain the ratio of the gravitational potentials and therefore the 
gravitational slip as
a function of model-independent observables 
\begin{equation}
\eta_{{\rm obs}}\equiv\frac{3P_{2}(1+z)^{3}}{2E^{2}\left(P_{3}+2+\frac{E'}{E}\right)}-1=\eta\,.\label{eq:etaobsp2p3}
\end{equation}
In order to distinguish the observables from the theoretical expectations,
we denoted the combination on the left-hand-side of this equation
as $\eta_{\mathrm{obs}}$. This is the quantity we will reconstruct using present data in a model-independent way. As advertised, $\eta_{obs}$ is independent of the initial power spectrum, of the bias, of the density of  matter, and of assumptions about the cosmic expansion (that is, we do not require a $\Lambda$CDM background or any other).

The parameter $P_{2}$ can be related to the  $E_g$ statistics, defined in the cosmological literature
(see \cite{Leonard2015} and references therein) as the expectation value of the ratio of lensing and galaxy clustering observables at a scale $k$
\begin{equation}
E_{g}=\bigg\langle\frac{a\nabla^{2}(\Psi-\Phi)}{3H_{0}^{2}f\delta}\bigg\rangle_{k}\,.
\end{equation}
Using the Poisson
equation (\ref{eq:poisson}) and the definition of the \textit{A, R, L, E}
variables (Eq.~\ref{eq:arle}), 
the relation with $P_{2}$ is simply given by 
\begin{equation}
P_{2}=2E_{g}\,.\label{eq.P2Eg}
\end{equation}
As we will mention in the next section, the available estimates of $E_g$ reduce to $P_2$ only under some conditions.
The $E_{g}$ statistics has been used several times as a test of modified
gravity (\cite{delaTorre:2016rxm}, \cite{Leonard2015}, \cite{Zhang2007}).
However, it is not {\it per se} a model-independent test. In fact, the theoretical
value of $E_{g}$ depends on $\Omega_{m0}$ and on $f$. Nevertheless,
$\Omega_{m0}$ is not an observable quantity. Distance indicators,
for instance Supernovae or BAO, measure $H(z)$ or its integral. To
estimate the matter fraction $\Omega_{m0}$ given $H(z)$, one needs
to assume that ``matter'' goes like $a^{-3}$ and the rest is parametrized
by an equation of state with few parameters. In modified gravity models,
neither is necessarily true. More in general, the ``dark degeneracy''
discussed for instance in \cite{Kunz2009} shows that the separation
between a matter component and a dark energy component is unavoidably
model dependent. There is a second problem with $E_{g}$, namely
the fact that the growth rate $f$ is estimated by solving the differential
equation of the perturbation growth. This requires initial conditions,
that are normally taken to be pure CDM at high redshift (this is for
instance how the well-known approximated formula $f\approx\Omega_{m}^{\gamma}(z)$
is obtained). Again, this assumption is not necessarily true in modified
gravity, as for example it is not true in the original Brans-Dicke
model. As a consequence of this, when we compare $E_{g}$ to the observed
value, we can never know whether any discrepancy with respect to $\Lambda$CDM and standard gravity
is due to a different value of $\Omega_{m0}$ or different initial
conditions, or it is a genuine signature of a non-standard modified gravity parameter $\Sigma$. So $E_g$ can be employed to test specific models, e.g. a $\Lambda$CDM expansion in standard gravity --  a very important task, indeed --  but not to measure
directy the properties of gravity.
In contrast, the statistics $\eta_{obs}$ of Eq.~(\ref{eq:etaobsp2p3})
is model-independent because it estimates directly $\eta$ without
any need to assume a value of $\Omega_{m0}$ nor to assume initial
conditions for $f$. Thus, if observationally one finds $\eta_{obs}\not=$1,
then $\Lambda$CDM and all the models in standard gravity and in which
dark matter is a  perfect fluid are ruled out. Finally,
we notice that in \cite{Amon2017} a cautionary remark is pointed
out, namely that their results about $E_{g}$ cannot be employed until
the tension between $\Omega_{m0}$ in different observational datasets
is resolved. This problem does not arise with $\eta_{obs}$.

Eq.~(\ref{eq:etaobsp2p3}) is a model-independent  estimate 
of $\eta$ that depends purely on observable quantities. As we already mentioned, the prefactor $h_{2}$ in Eq.~(\ref{eq:eta-quasistat}) is directly related to $c_{T}$, so
that for $c_{T}=c$, $h_{2}$ is equal unity. This means that
at large enough scales (for $k\to0$), $\eta\to1$ and one has the consistency relation
\begin{equation}
3P_{2}(1+z)^{3}=4E^{2}\left(P_{3}+2+\frac{E'}{E}\right) \;. \label{eq:p2ofp3}
\end{equation}
However, these large scales should still be sub-sound-horizon, so
that the quasi-static limit applies; moreover, the limit will actually
be different from unity in models without a mass scale, see e.g. \cite{Nersisyan:2018auj}.
Therefore, in practice, this large-scale consistency relation is not
particularly useful and we will not discuss it any longer.

\section{Data}

\label{datapart}

In this work we reconstruct $E(z)$, $P_{2}(z)$ and $P_{3}(z)$ using
the data listed in Tables \ref{tab.hzdata}, \ref{tab.covAlam}, \ref{tab.covWig},
\ref{tab.ezdata}, \ref{tab.egdata} and \ref{tab.fs8}, which are
also shown in figure \ref{plt.rawdata}. We use Hubble parameter data
to obtain $E(z)$ and $E'(z)$. For $P_{2}(z)$ we apply a simple
rescaling of $E_{g}(z)$ data, while $P_{3}(z)$ is reconstructed from
$f\sigma_{8}(z)$ and its derivative with respect to $\ln a$. We show in Table \ref{tab.fiducial}
the cosmological parameters from the TT+TE+EE+lowE+lensing Planck 2018 best-fits
\cite{Planck2018}, that we use to plot the $\Lambda\mathrm{CDM}$
curves of different cosmological functions in figure \ref{plt.rawdata}.
The details of the sources of the data will be explained below. \\

\begin{table}[h]
\centering %
\begin{tabular}{cccccc}
\hline 
$\Omega_{m0}$  & $\Omega_{DE}$  & $\Omega_{b}$ & $n_{s}$  & $\sigma_{8}$ & $H_{0}[\mathrm{km/s/Mpc}]$ \tabularnewline
\hline 
\hline 
0.3153  & 0.6847  & 0.0493  & 0.9649  & 0.8111 & 73.45 \tabularnewline
\hline 
\end{tabular}
\caption{Fiducial parameter values for our reference $\Lambda\mathrm{CDM}$ case,
using Planck 2018 data from TT+TE+EE+lowE+lensing \cite{Planck2018},
except for $H_{0}$, where we use the local value from the HST collaboration
\cite{2018ApJ...855..136R} as explained on the main text.}
\label{tab.fiducial} 
\end{table}

For the results of this work, we only use the $H_{0}$ value to normalize
$H(z)$ measurements into the dimensionless quantity $E(z)$. Notice
that $H_{0}$, contrary to $\Omega_{m0}$, is an observable quantity
that can be estimated from local kinematics in a way which is independent
of cosmology and modified gravity. Therefore, for the normalization
of the $E(z)$ measurements we need to choose a value of $H_0$, for instance 
from the recent results of the Planck collaboration \cite{Planck2018} 
or the value obtained by the HST collaboration \citep{2018ApJ...855..136R}. 

In this work, we choose the local value of $H_{0}$ determined by 
the HST collaboration \cite{2018ApJ...855..136R} 
which amounts to $H_{0}^{HST}=73.45\pm1.66\,[\textrm{km/s/Mpc}]$, 
because it is cosmology-independent. In Appendix \ref{app.H0} we discuss the results using the Planck value. 
Thus, by construction, we have an extra data point at $z=0$, namely,
$E(z=0)=1$. The uncertainty on $H_0$ propagates to all $E(z)$ values, 
and we take this into account as detailed in the next section.

\subsection{Hubble parameter data}\label{sec:hubble}

Regarding the Hubble parameter measurements, we have used the most
recent compilation of $H(z)$ data from \cite{Yu2017} (see Table
\ref{tab.hzdata}), including the measurements from \cite{Simon2004,Stern2009,Moresco:2012jh,Moresco:2015cya}, Baryon
Oscillation Spectroscopic Survey (BOSS) (\cite{Delubac:2014aqe,Font-Ribera:2013wce,Moresco2016}) and the Sloan Digital Sky Survey (SDSS) (\cite{Zhang2012},
\cite{Alam2016}).

In this compilation, the majority of the measurements was obtained
using the cosmic chronometric technique, labeled as method 1 in Table
\ref{tab.hzdata}. This method infers the expansion rate $dz/dt$
by taking the difference in redshift of a pair of passively-evolving
galaxies. The remaining measurements were obtained through the position
of the Baryon Acoustic Oscillation (BAO) peaks in the power spectrum
of a galaxy distribution for a given redshift. This is labeled as method 2 in Table
\ref{tab.hzdata}.

In addition to these, we use the recent results from \cite{Riess2017}
where a compilation of Supernovae Type Ia from CANDELS and the CLASH \textit{Multi-cycle
treasury program}  was analyzed,  providing six measurements
of the expansion rate $E(z)$, with considerably smaller error bars, compared to the other above mentioned techniques. 
These are listed on Table \ref{tab.ezdata}. 
In the original reference \cite{Riess2017}, the errors are not symmetric, therefore we recalculated symmetric errors, as the quadrature of the $1 \sigma$ bounds on the left and right side of the central value.

The measurements from \cite{Delubac:2014aqe} and \cite{Font-Ribera:2013wce}
are obtained using the BAO signal in the Lyman-$\alpha$ forest distribution
alone or cross correlated with Quasi-Stellar Objects (QSO) (for the
details of the observational methods, we refer the reader to the original papers).
Reference \citep{Alam2016} provides the covariance matrix of its
three $H(z)$ measurements obtained from the radial BAO galaxy distribution. We report the covariance matrix 
on Table \ref{tab.covAlam}. To this compilation we add
the results from the WiggleZ Dark Energy Survey \cite{Blake2012}
whose covariance matrix can be found on Table \ref{tab.covWig}.

The measurements of $H_0$ obtained with the cosmic chronometric technique are independent 
of large-scale cosmology and recent work \cite{Gomez-Valent2018} has shown that these data
prefer a lower value for the $H_{0}$ value. However, an upper
value can also be found if a different model for the processing of the galaxies spectra is chosen when using the data from \cite{Moresco2016}. For our fiducial results, we fix our choice of the Hubble parameter to the HST measurement.

As previously mentioned, the data points of the Hubble parameter $H(z)$ have to be
converted into the dimensionless expansion rate $E(z)$ by
dividing  by $H_{0}$, since we need $E(z)$, a model-independent observable.
For each measurement $H_i = H(z_i)$, we form
\begin{equation}
E_i = \frac{H_i}{H_0}
\end{equation} 
so that the error reads
\begin{equation}
\delta E_i = \frac{\delta H_i}{H_0} - H_i\frac{\delta H_0}{H^2_0} \, \, .
\end{equation} 
The covariance of this random matrix is then the expected value of the product of $\delta E_i$ and $\delta E_i$, which is
\begin{align} \label{H0cov}
\langle \delta E_i \delta E_j \rangle &= \langle \frac{\delta H_i}{H_0}
 \frac{\delta H_j}{H_0} \rangle + H_i H_j \langle \frac{\delta H_0}{H_0^2}
 \frac{\delta H_0}{H_0^2} \rangle \nonumber \\
 & = \frac{C^{(H)}_{ij}}{H_0^2} + E_i E_j \frac{(\sigma(H_0))^2}{H_0^2}  \,\,  ,
\end{align} 
where we have used the fact that errors on $H_0$ and $H_i$ are uncorrelated, therefore 
$\langle \delta H_0 \delta H_i \rangle = 0$, $C^{(H)}_{ij}$ is the covariance matrix of 
our data on the Hubble function $H(z)$ and $\sigma(H_0)$ is the error on $H_0$.
Equation \ref{H0cov} amounts to adding an extra covariance matrix to our standard data covariance matrix.

\subsection{$E_{g}$ data} \label{datapartEg}

We use the $E_{g}$ data compiled on Table \ref{tab.egdata}. This compilation
includes the results from KiDS+2dFLenS+GAMA \cite{Amon2017}, i.e,
a joint analysis of weak gravitational lensing, galaxy clustering and
redshift space distortions. We also include image and spectroscopic
measurements of the Red Cluster Sequence Lensing Survey (RCSLenS)
\cite{Blake2016} where the analysis combines the the Canada-France-Hawaii
Telescope Lensing Survey (CFHTLenS), the WiggleZ Dark Energy Survey
and the Baryon Oscillation Spectroscopic Survey (BOSS). Finally the
results of the VIMOS Public Extragalactic Redshift Survey (VIPERS) \cite{delaTorre:2016rxm}
is also accounted for in our data. They use redshift-space
distortions and galaxy-galaxy lensing.

These sources provide measurements in real space within the scales $ 3 < R_p < 60 h^{-1} $Mpc 
and in the linear
regime, which is the one we are interested in. They have been obtained
over a relatively narrow range of scales $\lambda$ meaning that we
can consider them relative to the $k=2\pi/\lambda$-th Fourier component,
as a first approximation. In any case, the discussion about the $k$-dependence
of $\eta$ is beyond the scope of this work, so the final result can
be seen as an average over the range of scales effectively employed
in the observations.
Moreover, in the estimation of $E_g$, based on \cite{Leonard2015}, one assumes that  the redshift of the lens galaxies can be approximated by a single value. With these approximations, indeed $E_g$ is equivalent to $P_2/2$, otherwise $E_g$ represents some sort of average value along the line of sight.
We caution that these approximations can have a systematic effect both on the measurement of $E_g$ and on our derivation of $\eta$. In a future work we will quantify the level of bias possibly introduced by these approximations in our estimate. For further details and discussion, see reference \cite{Blake2016} and \cite{delaTorre:2016rxm}.

\subsection{$f\sigma_{8}$ data}

In order to calculate the  variable $P_{3}$, we need to reconstruct 
$f\sigma_{8}(z)$ and its derivative as a function of redshift.
A compilation of the available data for $f\sigma_{8}(z)$ can be found in Table \ref{tab.fs8}.
This quantity can be obtained through measurements of the redshift-space
distortions (RSD) in the two point-correlation function of a galaxy survey.

Our data includes measurements from the 6dF Galaxy Survey \cite{Beutler2012},
the Subaru FMOS galaxy redshift survey (FastSound) \cite{Okumura2015},
WiggleZ Dark Energy Survey \cite{Blake2012}, VIMOS-VLT Deep Survey
(VVDS) \cite{Song2008}, VIMOS Public Extragalactic Redshift Survey
(VIPERS) \cite{delaTorre:2016rxm,Hawken2016,DelaTorre2013,Mohammad2017} and the Sloan Digital Sky Survey (SDSS) \cite{Howlett2015,Samushia2012, Tojeiro:2012rp, Chuang2012, Alam2016,Gil-Marin:2015sqa,Gil-Marin2016,Chuang2016}.
Other works in the literature which perform RSD measurements, but only report $f\sigma_{8}(z)$ values indirectly, such that we have to assume 
something on the bias or on the $\sigma_8$ relation, e.g. \cite{Cabre2009} and \cite{Guzzo2008}, will not be
considered for our purposes. 
Furthermore, for numerical reasons, before applying any reconstruction method, we will use these data taking the natural logarithm, i.e. $\ln f \sigma_8 (z) $, which allows us to compute the $P_3$ observable as a simple derivative with respect to $\ln a$.

\begin{figure}[H]
\includegraphics[width=0.325\textwidth]{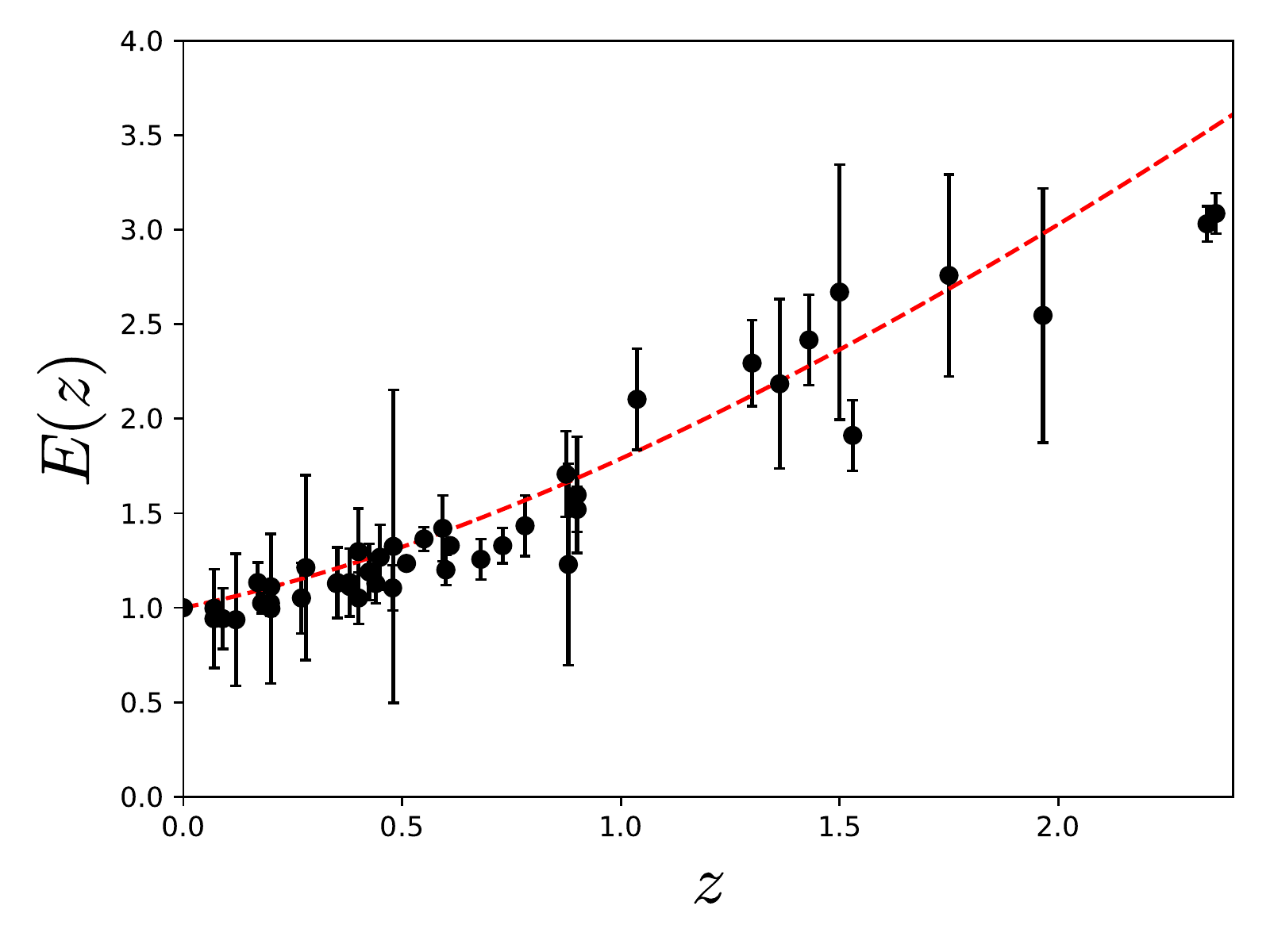}
\includegraphics[width=0.325\textwidth]{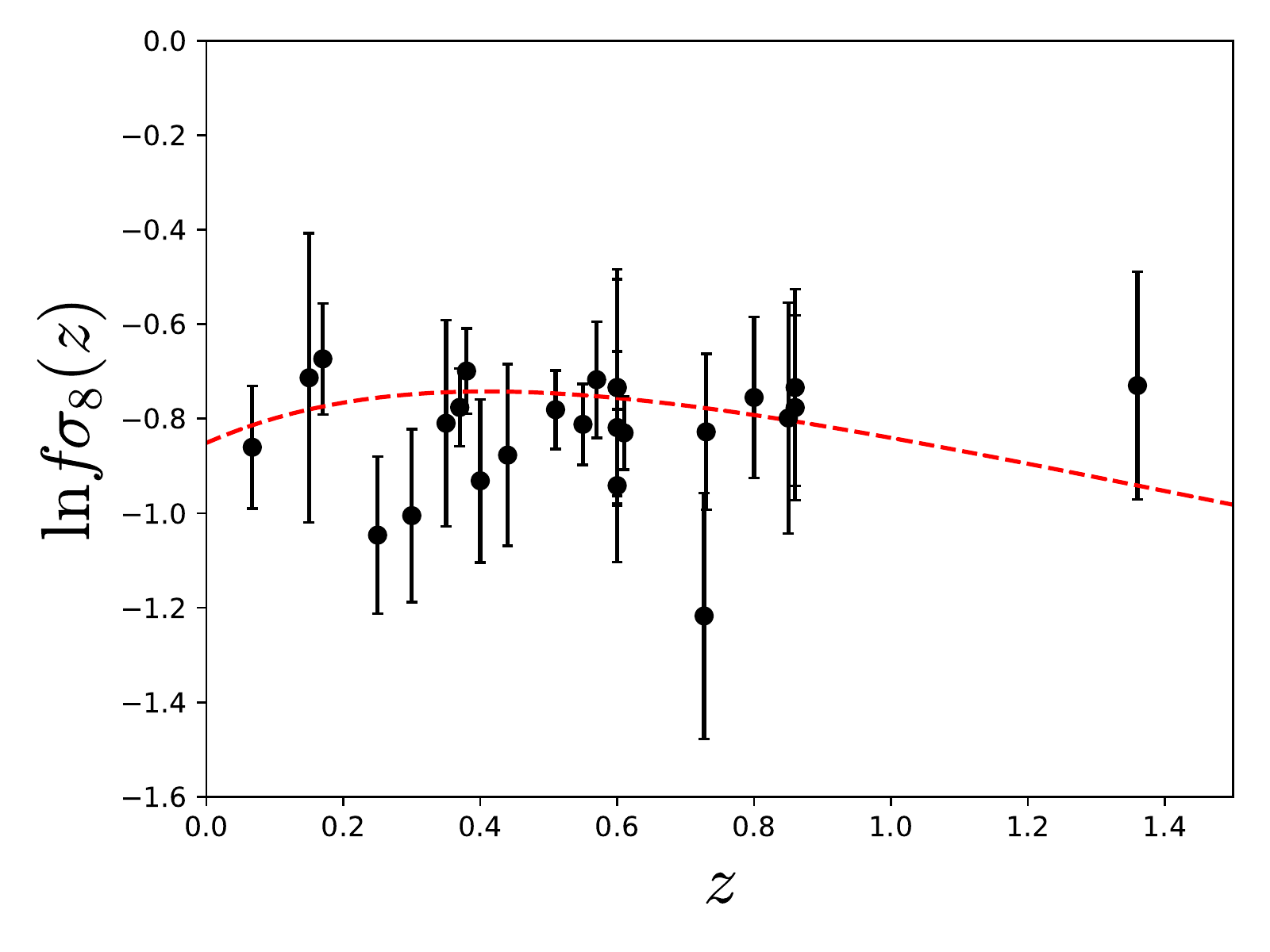}
\includegraphics[width=0.325\textwidth]{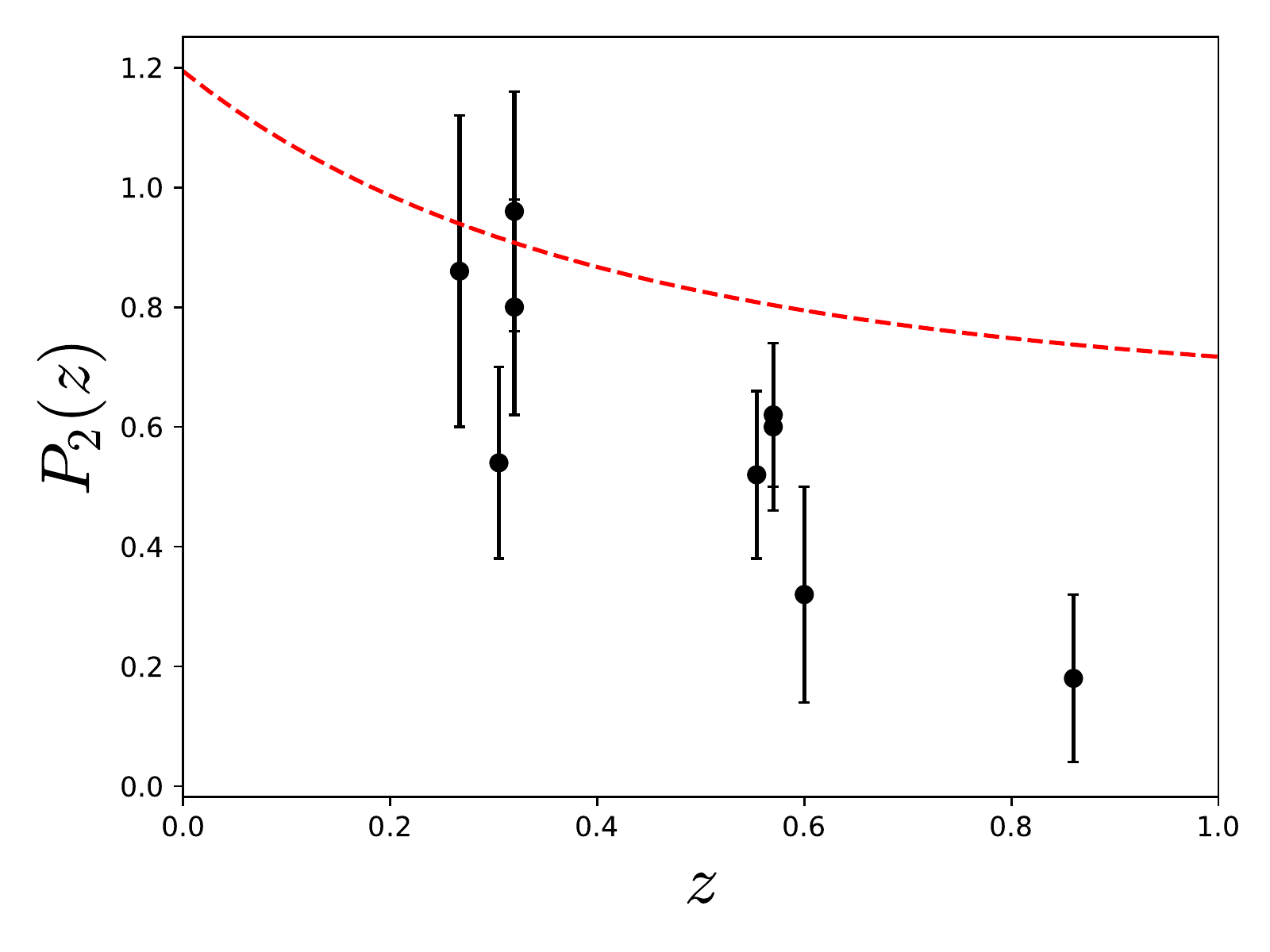}
\caption{Data sets used in this work (black dots with error bars), plotted
together with the corresponding reference $\Lambda\mathrm{CDM}$ prediction as
a function of redshift (solid red line), using a Planck 2018 cosmology
as reported on Table \ref{tab.fiducial}. \textbf{Left panel:}
$E(z)$ data from  Table \cite{Yu2017}. We used the value of $H_{0}$
from the HST collaboration to rescale part of the data points from
$H(z)$ to $E(z)$ (see main text). \textbf{Central panel:} Plot of
the natural logarithm of the $f\sigma_{8}$ data points from  Table \ref{tab.fs8}.
\textbf{Right panel:} Data set for $P_{2}$, obtained using $E_{g}$
data from Table \ref{tab.egdata}. For $z>0.5$ we see a large
discrepancy between $\Lambda\mathrm{CDM}$ and the data points, which was
also noted in \cite{Amon2017}.}
\label{plt.rawdata} 
\end{figure}

\section{Reconstruction of functions from data}

\label{recopart}

The main difficulties in obtaining $\eta_{\textrm{obs}}$ is that we need to take the ratios $P_2$ and $P_3$ at the same redshift, while we have datapoints at different redshifts, and that we need to take derivatives of $E(z)$ and $f\sigma_8(z)$. This essentially means
we need to have a reliable way to interpolate the data to reconstruct the underlying behavior.

There is no universally accepted method to interpolate data. Depending on how many assumptions one makes regarding the theoretical model, e.g. whether the reconstructed functions need just to be continuous, or smooth, depending on few or many parameters, etc., one gets unavoidably different results, especially in the final errors.
Here, we consider and compare three methods to obtain the value of $\eta_{\rm obs}$: binning, Gaussian Process, and generalized polynomial
regression. 

\subsection{Binning}

One intuitive way to perform data reconstruction is to assemble the
data into bins. This consists in dividing the data into particular
redshift intervals and, for each of these intervals (bins),  calculating
the average  of the data contained in that bin.
Denoting $s_k \equiv s(z_{k})$ as a generic data value, with dependent variable
$s$, located at the point $z_{k}$ with error $\sigma_{k}^{s}$,
the binning procedure is done by applying the following formula 
\begin{equation}
\bar{s}_{i}=\frac{\sum_{k}^{N_{i}}s_{k}\big(\sigma_{k}^{s}\big)^{-2}}{\sum_{k}^{N_{i}}\big(\sigma_{k}^{s}\big)^{-2}},\qquad\sigma_{i}^{\bar{s}}=\frac{1}{\sqrt{\sum_{k}^{N_{i}}\big(\sigma_{k}^{s}\big)^{-2}}}\quad.
\end{equation}
where $N_{i}$ is the number of data points inside the bin $i$, $\bar{s}_{i}$
is the new value of the dependent variable at the center of the bin
$z_{i}$, where $z_{i}=(z_{k+1}-z_{k})/2$ is simply the arithmetic mean between the upper and lower borders
of the bin. The new error at this point is $\sigma_{i}^{\bar{s}}$. 
This means that we are converting the information of the
subset of data contained in a specific bin into one unique data point
by taking the weighted average for the data values and the data errors over
all points contained in that interval.  The square of the new error bar
at the center of the bin, namely $(\sigma_{i}^{\bar{s}})^{2}$, is
then the mean of the errors squared from all the $N_{i}$ points contained
in the bin with index $i$.

To reconstruct our main observable $\eta_{\rm obs}$, we also need to
compute the derivatives of the data for the functions $E(z)$ and
$\ln (f\sigma_{8}(z))$, at the exact same redshifts as for the other
functions. Therefore, we need to bin the original data in alternative
bins centered at new points $z_{j}$, so that using finite differences
we can compute the derivative of the dependent variable and its associated
error at the $z_{i}$ in the following form 
\begin{equation}
\bar{s}'_{i}=-(1+z_{i})\frac{\bar{s}(z_{j+1})-\bar{s}(z_{j})}{\Delta z_{j}},\qquad\sigma_{i}^{\bar{s}'}=(1+z_{i})\frac{1}{\Delta z_{j}}\sqrt{(\sigma_{j+1}^{\bar{s}})^{2}+(\sigma_{j}^{\bar{s}})^{2}}\quad.
\end{equation}
where $\Delta z_{j}=z_{j+1}-z_{j}$ and remembering that a prime denotes
a derivative with respect to $\ln a$.

Our observable $\eta_{\rm obs}$ is estimated as in Eq.~(\ref{eq:etaobsp2p3})
through $E(z)$, $P_{2}$, $P_{3}$ and $E'(z)$, which we will denote
generally as $y^{(1)}$, $y^{(2)}$, $y^{(3)}$ and $y^{(4)}$, respectively.
Consequently, to calculate the final error on $\eta_{\rm obs}$, we use
standard error propagation, assuming no correlation among the $y^{(i)}$
variables, so that the error $\sigma_{i}^{\eta_{\rm obs}}$ at the redshift
$z_{i}$ is specifically 
\begin{equation}
(\sigma_{i}^{\eta_{\rm obs}})^2=\sum_{\alpha=1}^{4}\bigg(\sigma_{i}^{y^{(\alpha)}}\frac{\partial\eta_{\rm obs}(z_{i})}{\partial y^{(\alpha)}}\bigg)^{2}
\end{equation}
where we also assume that the bins are large enough, such that the
correlation among the bins is negligible. In this way, Eq.~(\ref{eq:etaobsp2p3})
and its estimated error can be evaluated at the centers of the bins
$z_{i}$. However, the maximum number of final bins $N_{i}$ is constrained
by the number of data points available for the smallest data set among
the $y^{(\alpha)}$ functions.
We will present results on
the binning method with
more detail in section \ref{resupart}.

\subsection{Gaussian Process}

Another way of reconstructing a continuous function from a dataset is
using the method of Gaussian Process (see \cite{GPbook} for a comprehensive description).
A Gaussian Process (GP) can be regarded as the generalization of Gaussian distributions
to the space of functions, since it provides a probability distribution over continuous functions
instead of a distribution over a random variable. Considering a dataset
$\mathcal{D}=\{(x_{i},y_{i})\vert i=1,...n\}$ of \textit{n} observables
where $x_{i}$ are deterministic variables and $y_{i}$ random variables,
the goal is to obtain a continuous function $f(x)$ that best describes
the dataset. A function $f$ evaluated at a point $x$ is a Gaussian
random variable with mean $\mu$ and variance $\mathrm{Var(f)}$.
The $f(x)$ values depend on the function value evaluated at another point
$x'$. The relation between the value of the function at these two points 
can be given by a covariance function $\mathrm{cov}(f(x),f(x'))=k(x,x')$, 
which evaluated at $x = x'$ gives the variance $\mathrm{Var}(f(x))=k(x,x)$.
So, the distribution of functions at the point $x$ is characterized by (for more details, see \cite{Seikel2012})
\begin{equation}
\mu(f(x))=\mathcal{E}[f(x)]\qquad k(x,x')=\mathcal{E}[(f(x)-\mu(x))(f(x')-\mu(x'))] \;,
\end{equation}
where $\mathcal{E}$  is the expected value.

The covariance function $k(x,x')$ is in principle arbitrary. Since
we are interested in reconstructing the derivative of the data, we need to chose a differentiable function.
A Gaussian covariance function
\begin{equation}
k(x,x')=\sigma_{f}^{2}\exp\bigg[-\frac{(x-x')^{2}}{2\ell_{f}^{2}}\bigg] 
\end{equation}
is the covariance function that we choose in this work, as it is the most common and it has the least number of parameters. 
In the results section \ref{resupart}, we will discuss  how this assumption does not change considerably our results.
This function depends on the hyperparameters $\sigma_{f}$ and $\ell_{f}$, that allow to
set the shape of the covariance function, which acts as a form of prior on the set of possible functions that we can obtain with the GP method. 
The hyperparameter $\ell_{f}$ can be considered as the typical correlation length scale of the independent variable,
while the signal variance $\sigma_{f}$, can be thought of as the typical variation scale of the dependent variable.

In a Gaussian Process using real data $(x_{i},y_{i})$ where $y_i = f(x_i) + \epsilon_i$, the errors are assumed to be Gaussian 
and the observations to be scattered around the underlying function.
The noise $\epsilon_{i}$ is Gaussian with covariance matrix $C$, which needs to be taken into account for the joint likelihood function.
This means that the reconstruction itself depends on the number and quality of data available.

Following a Bayesian approach, one can compute the joint likelihood function
for the data and the reconstructed function. 
Thus, for a Gaussian prior for both the data and the random functions, one can marginalize over the space of functions $f$ and obtain the logarithm of the marginal likelihood as (see \cite{Seikel2012})
\begin{align}
\begin{split}
\ln\mathcal{L}=&-\frac{1}{2}\sum_{i,j=1}^{N}\Bigg\{\big[y_{i}-\mu(x_{i})\big][k(x_i,x_j)+C(x_i,x_j)]^{-1}\big[y_j-\mu(x_{j})\big]\Bigg\} \\
&-\frac{1}{2} \ln \Big\vert k(x_i,x_j)+C(x_i,x_j) \Big\vert - \frac{N}{2}\ln2\pi \ . \label{eq:marglike}
\end{split}
\end{align}

Maximizing the logarithm of the marginal likelihood gives then the optimal hyperparameters $\sigma_{f}$ and $\ell_{f}$.
In a full Bayesian approach, one should marginalize over the hyperparameters, using Monte Carlo Markov chain (MCMC) algorithms, 
in order to obtain the fully marginalized posterior distribution on the reconstructed function.
As suggested in \cite{Seikel2012}, we assume that the probability distribution of the hyperparameters is sharply peaked, which allows us to take them out of the integration and effectively fix them to
their optimal values.

The Gaussian Process algorithm is implemented in a publicly available python code, 
named GaPP (Seikel et al. (2012) \cite{Seikel2012}).
The GaPP code computes the continuous function of a given dataset and its derivatives up to third order,
for a multi-dimensional dataset. It also takes into account
correlated errors in the data and allows one to
choose among different covariance functions, also known as kernel
functions. 
For the case of the Gaussian kernel function as described above, 
the $\sigma_{f}$ and $\ell_{f}$ parameters are optimized
by the GaPP code through the maximization of the logarithm of the marginal
likelihood function in Eq.~(\ref{eq:marglike}).

Also, for the case of reconstructing the derivative of the data, a
covariance between the reconstruction of $f$ and $f'$ arises, that
should also be determined by a Monte Carlo sampling. GaPP takes a
first order approximation and uses statistical error propagation which
is valid for small errors. 
These approximations may have an impact on the final constraints of
this work, particularly as underestimated errors on the reconstructed
function as discussed on the original reference \cite{Seikel2012}.

For each of the data sets, we will use the GaPP code to reconstruct the underlying function and its derivative 
where we did not specify any prior on the hyperparameters or the mean function of the Gaussian Process to remain agnostic towards these choices. The details of our approach using this code concerning the chosen hyperparameters and covariance functions will be discussed in Section \ref{resupart}.

\subsection{Polynomial regression}

\label{sec:LRG}

As a third reconstruction method, we use a generalized polynomial 
regression, a widely used method to obtain model parameters from data. 
Since we want to do this as model-independently as possible, we do
not impose a priori any polynomial order for the reconstruction, but
we let the data decide which is the maximum possible order. 
In the following we will describe the standard method of polynomial regression. 
Nevertheless, there are a
number of complications due to the application to differentiated data with correlated errors,
so that we will discuss the method in detail in App. \ref{app.LRG}.

We start by assuming that we have
$N$ data points $y_{i}$, one for each value of the \emph{independent} variable
$x_{i}$ (which are \emph{not} random variables) and that 
\begin{equation}
y_{i}=f_{i}+e_{i}
\end{equation}
where $e_{i}$ are errors (random variables) which are assumed to
be distributed as Gaussian variables. Here $f_{i}$ are theoretical
functions that depend linearly on a number of parameters $A_{\alpha}$
\begin{equation}
f_{i}=\sum_{\alpha}\bar{A}_{\alpha}g_{i\alpha}\label{eq:linearfunc}
\end{equation}
where $g_{i\alpha}(x_{i})$ are functions of the variable $x_{i}$. This is the definition of a linear model.
Defining the matrix of basis functions as $G$ and the
data vector as $D$ in the following way 
\begin{eqnarray}
G_{\alpha\beta} & \equiv & g_{\beta i}C_{ij}^{-1}g_{\alpha j}\\
D_{\alpha} & \equiv & y_{i}C_{ij}^{-1}g_{\alpha j}
\end{eqnarray}
(always summing over repeated Latin indexes), where $C_{ij}$ is the
data covariance matrix, we can see that the linear model
can be written as
\begin{equation}
\mathbf{G}\mathbf{A}=\mathbf{D}\ .\label{eq:normaleqs}
\end{equation}
We are interested in finding the coefficients $\boldsymbol{A}=\{A_{0},A_{1},...\}$
of the model. To do so, we can invert the above equation to solve for $\boldsymbol{A}$
as 
\begin{equation}
\mathbf{\bar{A}}=\mathbf{G}^{-1}\mathbf{D}\ ,\label{eq:normalsol}
\end{equation}
which is also known as the \textit{normal equation}.

If the prior is uniform in an infinite range (improper prior), the
parameters in the linear problem have a Gaussian posterior with mean
$\bar{\mathbf{A}}$ and correlation matrix given by the inverse of
its Fisher matrix. Since in the linear problem the data covariance
matrix does not depend on the parameters, we have the following Fisher
matrix 
\begin{equation}
F_{\alpha\beta}\equiv C_{ij}^{-1}\frac{\partial f_{i}}{\partial\bar{A}_{\alpha}}\frac{\partial f_{j}}{\partial\bar{A}_{\beta}}=C_{ij}^{-1}g_{\alpha i}g_{\beta j}=G_{\alpha\beta} \;.\label{eq:fullfm-1}
\end{equation}
Once the coefficients are known, we can obtain the data
values on a point $x_{A}$, which is not one of the points present
in the data, using the expression in Eq.~(\ref{eq:linearfunc}) and
evaluating it at $x_{A}$, namely $f_{A}=\sum_{\alpha}\bar{A}_{\alpha}g_{A\alpha}$,
where $g_{A\alpha}$ means the function $g_{\alpha}$ evaluated at
$x_{A}$, with an error $\sigma_{A}^{2}=F_{\alpha\beta}^{-1}g_{A\alpha}g_{A\beta}$.
We can select a number of arbitrary points $x_{A,B,C}$ and obtain
the error matrix for the reconstructed function at these points as
\begin{equation}
C_{AB}=F_{\alpha\beta}^{-1}g_{A\alpha}g_{B\beta} \;.
\end{equation}

In our particular case,
we have three datasets $(y^{(0)},y^{(1)},y^{(2)})=(\ln(fs_{8}(z)),E(z),E_{G}(z))$
and we wish to estimate the error on a function $\eta_{{\rm obs}}(y^{(1)},y^{(2)},y^{(3)},y^{(4)})$,
where $y^{(4)}=y^{(1)'}$and $y^{(3)}=y^{(0)'}$ where a prime denotes, as already mentioned, a derivative with respect to $\ln(a)$.
We leave the details for App.\ref{app.LRG}. 
The only issue we discuss here is the order of the polynomial. The
order is of course in principle arbitrary, up to the number of data
points for each data set. However, it is clear that with too many free parameters the
resulting $\chi^{2}$ will be very close to zero, that is statistically
unlikely. At the same time, too many parameters also render the numerical
Fisher matrix computationally unstable (producing, e.g., a non-positive
definite matrix) and the polynomial wildly oscillating. On the other
hand, too few parameters restrict the allowed family of functions.
Therefore, we select the order of the polynomial function by choosing
the polynomial degree for which the reduced chi-squared $\chi_{red}^{2}=\chi^{2}/(N-P)$,
is closest to 1 and such that the Fisher matrix is positive definite. 
Since our datasets contains data points from different experiments,
there are some data points located at the same redshift or very close
to each other, with different values of the dependent variable. In
the case of a perfect fit, the polynomial would go through all points
leading to spurious oscillations. For this reason, we take the weighted
average of data points that are closer than $\Delta z=0.01$ in redshift,
before using them as an input into the polynomial regression algorithm. 

\begin{figure}[htp]
\centering
\includegraphics[width=0.45\textwidth]{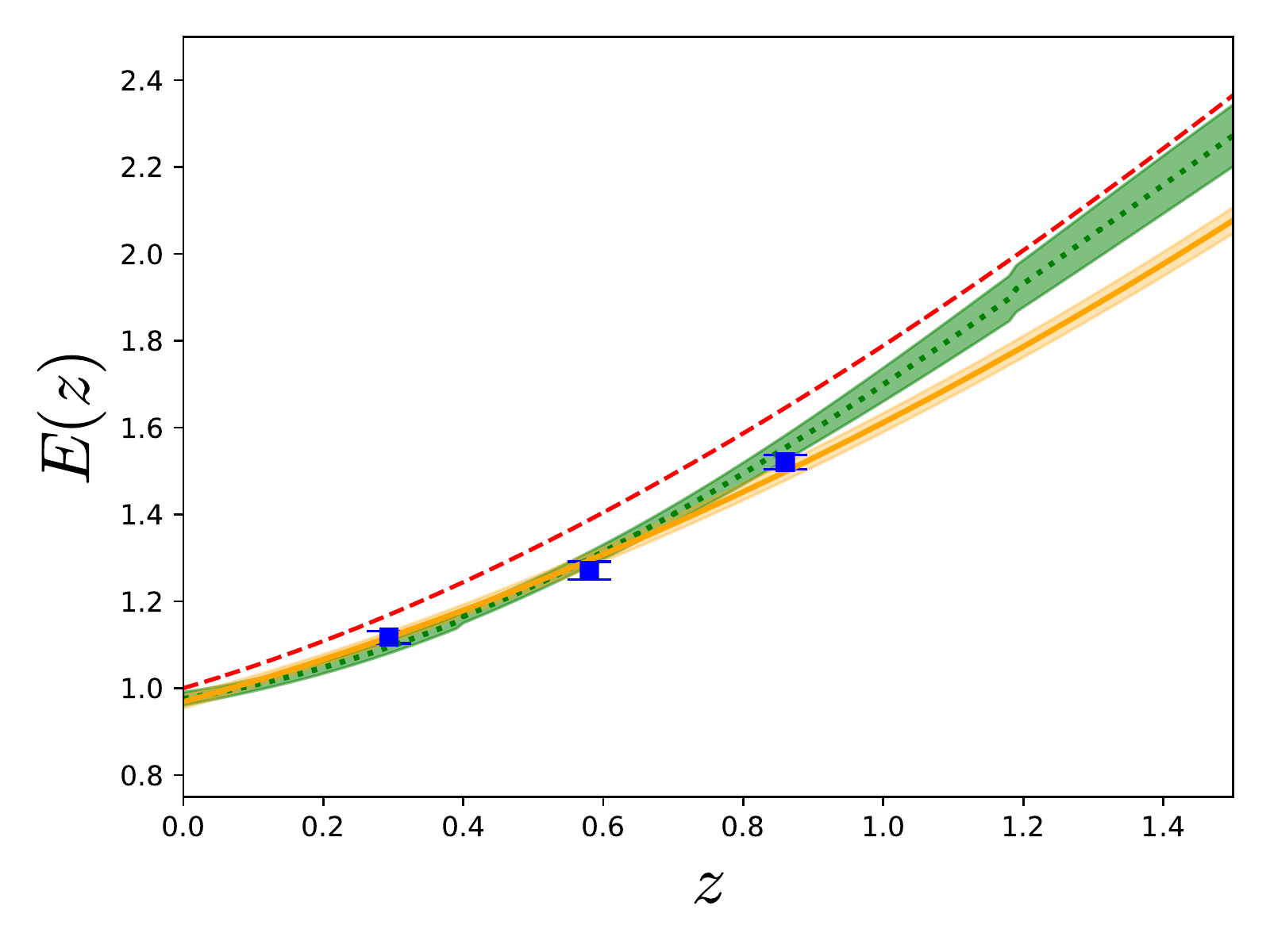}
\includegraphics[width=0.45\textwidth]{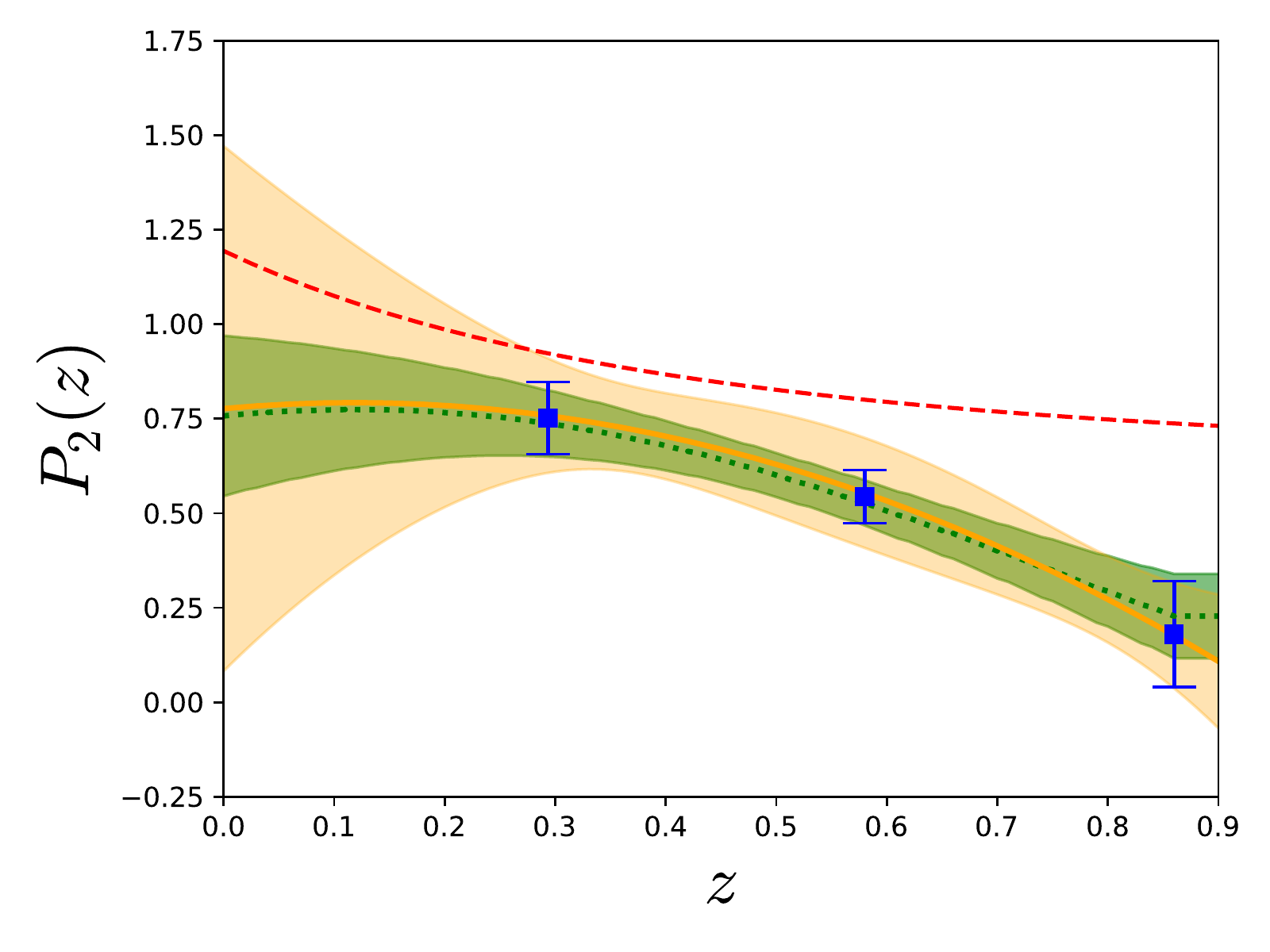}\\
\includegraphics[width=0.45\textwidth]{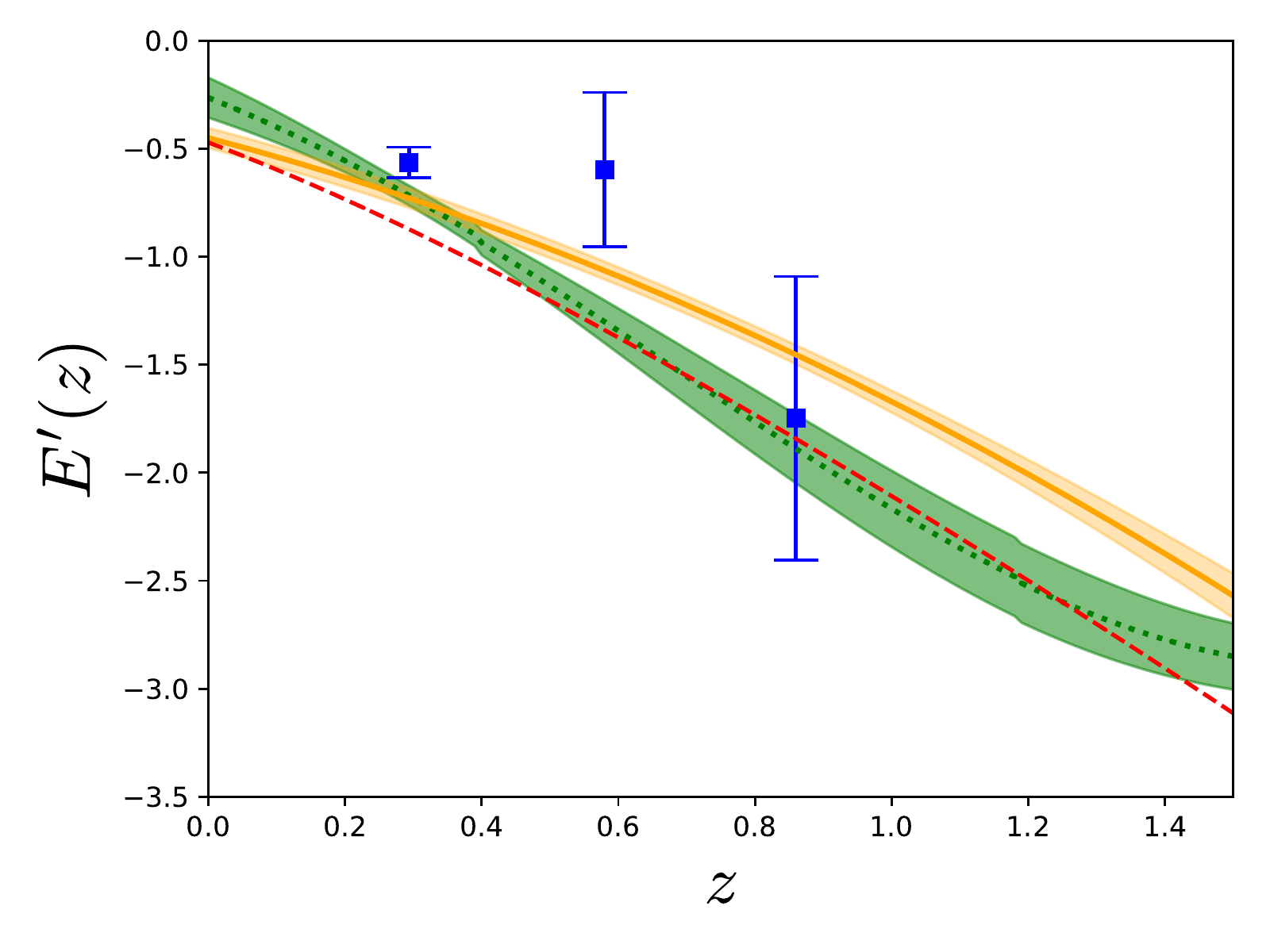}
\includegraphics[width=0.45\textwidth]{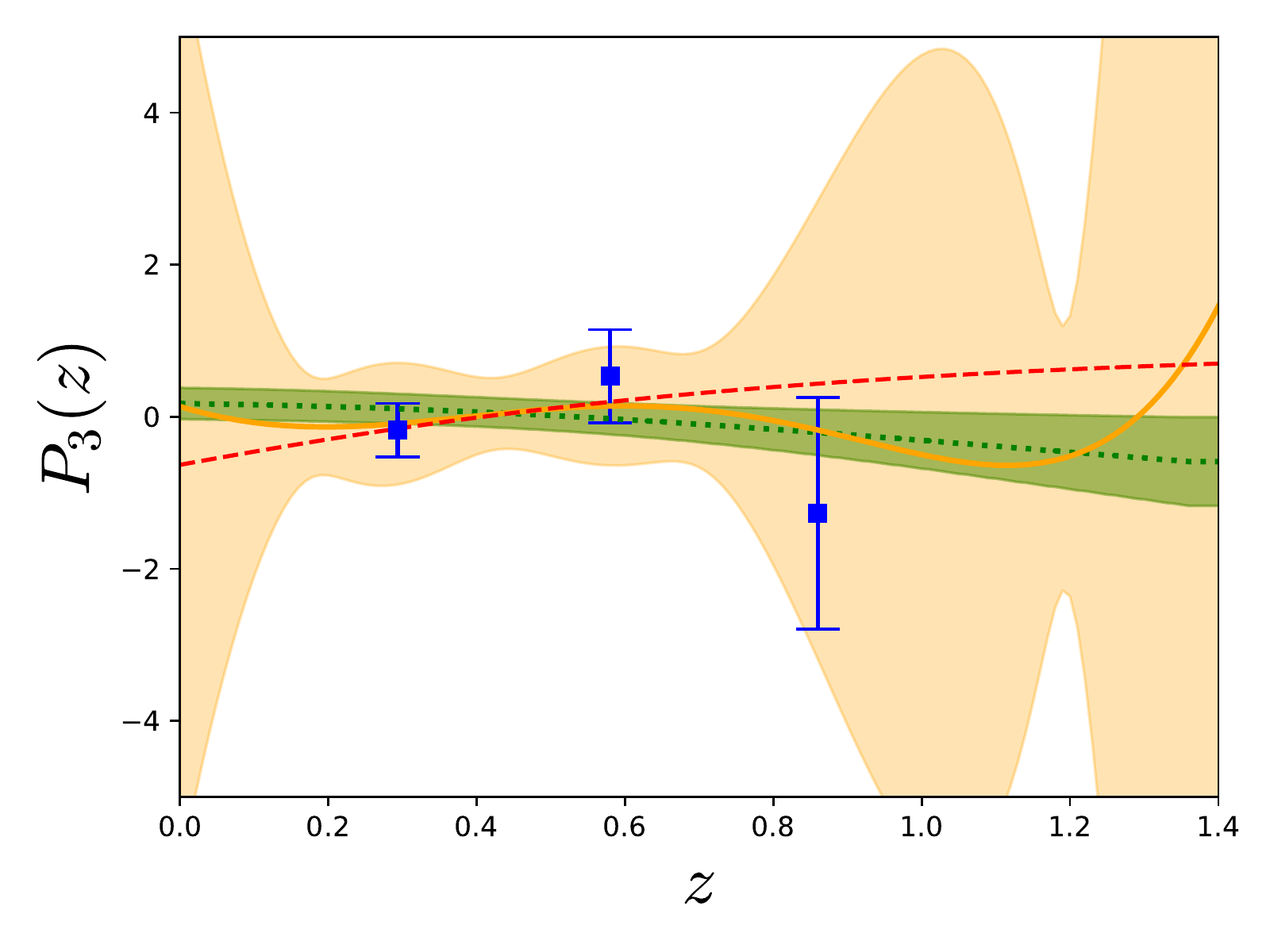}
\caption{Comparison of the three reconstruction methods for each of the model-independent variables. The binning method in blue squares
with error bars, Gaussian Process as a green dotted line with green bands, polynomial regression as a solid yellow line with yellow bands. All of them depicting the $1\sigma$ uncertainty.
\textbf{Left panel:} Plot of the reconstructed $E(z)$ function on
the top and its derivative $E'(z)$ on the bottom.
\textbf{Right panel:} Plot of the reconstructed $P_{2}(z)$ function
on the top and the reconstructed $P_3(z)$ function on
the bottom. For each case, we show the theoretical prediction of our reference $\Lambda\mathrm{CDM}$ model
as a red dashed curve.}
\label{plt.rec} 
\end{figure}

\section{Results}

\label{resupart}

Let us now discuss the results of the final observable $\eta_{{\rm obs}}$
for each of these methods.
The binning method contains the least number of assumptions compared to the polynomial regression
or the Gaussian Process method. 
It is essentially a weighted average over the data points and its
error bars at each redshift bin. Since we need to take
derivatives in order to calculate $P_{3}$ and $E'$, and we
have few data points, we opt to compute finite difference derivatives.
This has the caveat that it introduces correlations among the errors
of the function and its derivatives, that we cannot take into account
with this simple method. Moreover, for the binning method, we do not take into account possible
non-diagonal covariance matrices for the data, which we do for polynomial regression and the Gaussian Process reconstruction.

Figure \ref{plt.rec} shows the
reconstructed functions obtained by the binning method, the Gaussian Process  and with polynomial regression,
alongside with the theoretical prediction of the standard $\Lambda$CDM
model. In all cases the error bars or the bands
represent the $1\sigma$ uncertainty.

With the binning method, the number of bins is limited by the maximum
number of existing data redshifts from the smallest data set corresponding
to one of our model-independent observables. In this case, this is
the quantity $E_{g}$, for which we have effectively only three redshift
bins. Looking at Table \ref{tab.egdata} and comparing with Figure \ref{plt.rawdata}, 
we can see that there are
nine $E_g$ data points, but most of them are very close to each other in redshift, due to being
measured by different collaborations or at different scales in real
space for the same $z$. As explained in the data section above, we just regard this
data as an average over different scales, assuming that non-linear
corrections have been correctly taken into account by the respective
experimental collaboration. Since we do not have to take derivatives
of $E_{g}$, or equivalently $P_{2}$, this leaves us with three possible
redshift bins, centered at $z_{1}=0.294$, $z_{2}=0.580$ and $z_{3}=0.860$, all of them with
an approximate bin width of $\Delta z \approx 0.29$.
At these redshifts we obtain $\eta_{{\rm obs}}(z_{1})=0.48 \pm 0.45$,
$\eta_{{\rm obs}}(z_{2})=-0.03 \pm 0.34$ and $\eta_{{\rm obs}}(z_{3})=-2.78 \pm 6.84$.
These values and the estimation of the intermediate model-independent
quantities can be seen in Table \ref{tab.results}.

Regarding the Gaussian Process method, we have computed the normalized Hubble function and its derivative,
$E(z)$ and $E'(z)$ with the \textit{dgp} module of the GaPP code.
Using the data of Table \ref{tab.ezdata} and its correlation matrix,
we reconstructed the $E(z)$ and $E'(z)$ for the redshift interval
of the data using the Gaussian function as the covariance function and initial
values of the hyperparameters $\theta=[\sigma_{f}=0.5,\ell_{f}=0.5]$ that later are estimated
by the code. The same procedure was done for the $P_{2}(z)$ data, obtained by Eq.~(\ref{eq.P2Eg}) using the Table \ref{tab.egdata}. 
We obtain for $E(z)$ and $E'(z)$ functions the hyperparameters $\sigma_{f}=2.12$ and $\ell_{f}=2.06$ and
for the $P_{2}$ function, $\sigma_{f}=0.58$ and $\ell_{f}=0.67$.

For the $P_3(z)$ observable, the hyperparameters obtained by the GaPP code led to 
a very flat and unrealistic reconstruction, that suggested us to take another approach for obtaining the optimal hyperparameters. More details can be found on Appendix \ref{app.GP}. 
We sampled the logarithm of the marginal likelihood on a grid of hyperparameters $\sigma_{f}$, $\ell_{f}$ 
from 0.01 to 2, setting this way a prior with the redshift range of the dataset,
and  300 points equally separated in log-space for each dimension. 
Remember that the hyperparameter $\ell_{f}$ constrains the typical scale on the independent variable $z$. 
Thus, as an additional prior, we impose that $\ell_{f}$ needs to be smaller than the redshift range of the data, which was not guaranteed by the default GaPP code.
Then we chose the pair of hyperparameters corresponding to the maximum
of the log-marginal likehood Eq.~(\ref{eq:marglike}).
Therefore, for the $\ln (f\sigma_{8}(z))$ data, we obtain $\sigma_{f}=0.549$ and $\ell_{f}=1.361$.
Its reconstructed derivative $P_3$ can be seen in the lower right panel of Figure \ref{plt.rec}.
The function remains relatively flat, compared to the one given by other methods, but this approach has improved the 
determination of this observable, as further justified in Appendix \ref{app.GP}. \\

Regarding the choice of the kernel function, several functions
were compared, each of them with a different number of parameters to see the impact
on the output. We tested the Gaussian kernel with two parameters, ($\sigma_{f},\ell_f$);
the rational quadratic kernel with three parameters and the double Gaussian
kernel with four parameters (see the original reference for the explicit implemented
formula \cite{Seikel2012}). We performed tests using the $H(z)$
data obtained with the cosmic chronometer technique and the $f\sigma_{8}(z)$
data. Our tests show that the different choices shift the reconstructed function
up to $6\%$ on its central value compared
to the Gaussian kernel function. This happens for $H(z)$ while the effect is negligible
for $f\sigma_{8}(z)$. 
Taking into account the above choices and procedure, we report that with the Gaussian Process method 
we obtain $\eta_{{\rm obs}}(z_{1})=  0.38 \pm 0.23 $,
$\eta_{{\rm obs}}(z_{2})=0.91 \pm 0.36 $ and $\eta_{{\rm obs}}(z_{3})=0.58 \pm 0.93$. 

\input{allfinaldata.tex}

For the polynomial regression method, we find $\eta_{{\rm obs}}(z_{1})=0.57\pm1.05$,
$\eta_{{\rm obs}}(z_{2})=0.48\pm0.96$ and $\eta_{{\rm obs}}(z_{3})=-0.11\pm3.21$.
Note that we applied the criteria of a $\chi_{red}^{2}$ closest to
one and a positive definite Fisher matrix to chose the order of the
polynomial for each of the datasets. These criteria led to a choice
of a polynomial of order 3 for the $E(z)$ and $E_{g}(z)$ data and
order 6 for the $\ln (f\sigma_{8}(z))$ data. These polynomials can be seen in 
Figure \ref{plt.rec} as solid yellow lines, together with their $1\sigma$ uncertainty bands.
The higher order of the polynomial of  $\ln (f\sigma_{8}(z))$ explains the "bumpiness" 
of the reconstruction of $P_3$, leading to larger errors on this observable in comparison to the GP method.

\begin{figure}[htbp]
\centering
\includegraphics[width=0.7\textwidth]{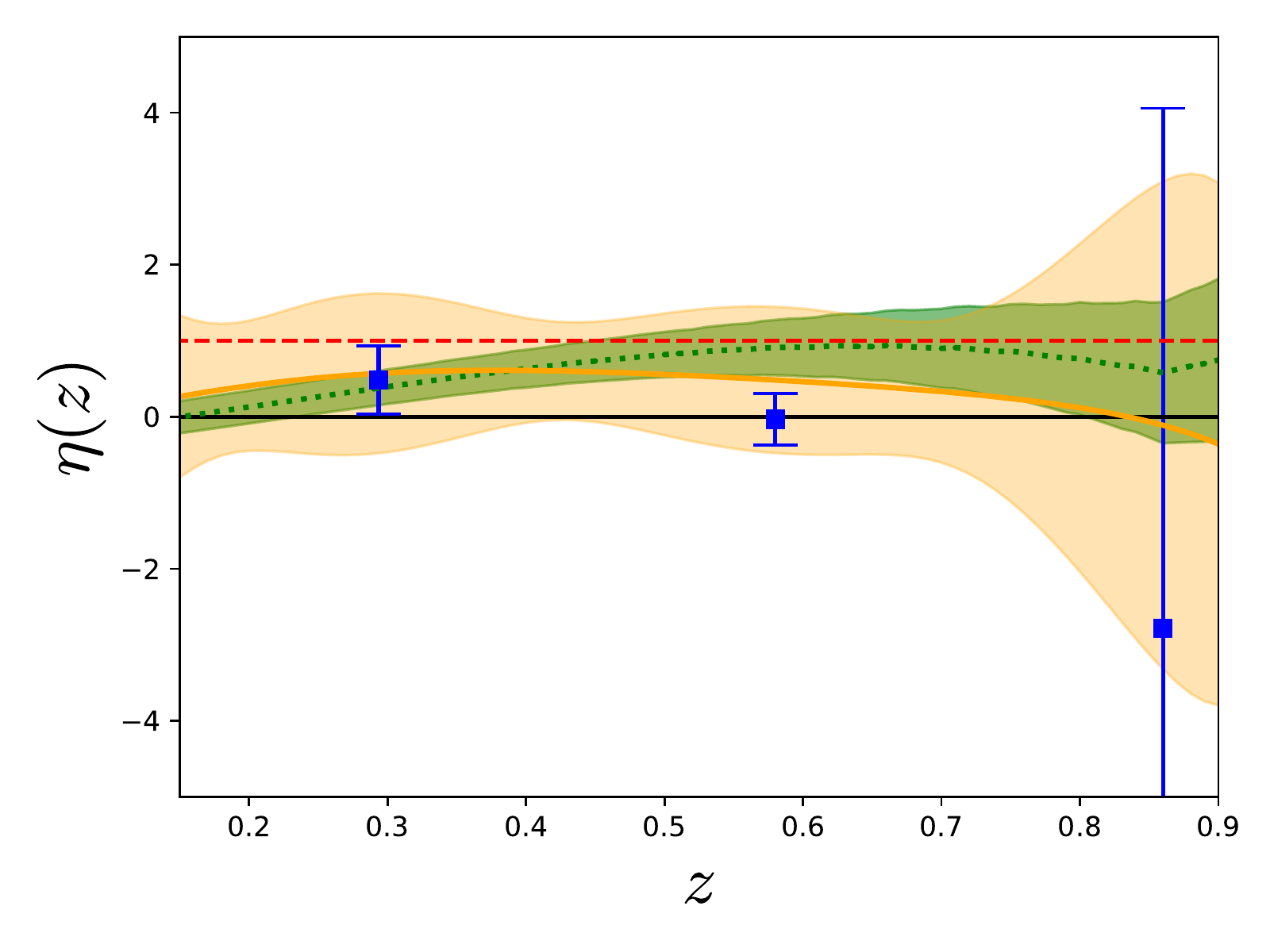}
\caption{Plot of the reconstructed $\eta_{\rm obs}$ as a function of redshift,
using the binning method (blue squares), Gaussian Process (green
dotted line) and polynomial regression (yellow solid line). The corresponding
error bands (error bars for the binning method), represent the $1\sigma$
estimated error on the reconstruction. As a reference, we show in
a dashed red line the value in  standard gravity.}
\label{plt.etabar-all} 
\end{figure}

In Fig. \ref{plt.etabar-all} we show the reconstructed $\eta_{\rm obs}$
as a function of redshift with the three different methods, again with GP in a green dashed line,
polynomial regression in a yellow solid line and the binning method in blue squares with error bars. 
It is possible to
conclude that the methods are consistent with each other, within their  $1\sigma$ uncertainties and that
in most bins the results are consistent with the standard gravity scenario.
We find that the error bars of the Gaussian Process reconstruction are generally
smaller than the other methods, such that at the lowest redshift, 
GP is not compatible with $\eta_{\rm obs}=1$ at nearly $2\sigma$, 
while in the case of the binning method at the intermediate redshift, $z=0.58$, the tension is nearly $3\sigma$.

As detailed in section \ref{sec:hubble}, we need to choose a value of $H_0$ to obtain the dimensionless Hubble function $E(z)$. We tested that our results do not change significantly with a different choice of $H_0$.  The comparison between the value from the Planck 2018 collaboration and the HST collaboration is described on Appendix \ref{app.H0}.

Finally, we can combine the estimates at three redshifts
of Table \ref{tab.results} into a single value. Assuming a constant
$\eta_{\rm obs}$ in this entire observed range and performing a simple
weighted average, we find finally $\eta_{\rm obs}=0.15 \pm 0.27$ (binning),
$\eta_{\rm obs}=0.53 \pm 0.19$ (Gaussian Process) and $\eta_{\rm obs}=0.49 \pm 0.69$
(polynomial regression).
The Gaussian Process method yields the smallest error and would exclude standard gravity. 
However, despite being sometimes advertised as ``model-independent'', we believe that 
this method actually makes a strong assumption, since it compresses the ignorance
about the reconstruction into a kernel function that depends on two or a small number of parameters, 
which are often not even fully marginalized over, as we did in our case. 
Also the binning method taken at face value would rule out standard gravity. However, as already mentioned, 
we did not take into account the correlation induced by the finite differences, and this might have decreased
the overall error.
Overall, we think the polynomial regression method is the most satisfactory one, 
providing the best compromise between the least number of assumptions and the best 
estimation of the data derivative. Therefore, we consider it as our ``fiducial'' result.

\section{Conclusions}

Large scale surveys like Euclid will soon allow to
combine lensing and clustering data of unprecedented quality and quantity
to probe gravity. To this aim, it is important to perform both null-tests
of specific models, like $\Lambda$CDM, and to measure the properties
of modified gravity in a way that does not depend on too many assumptions.

One of the clearest way to test gravity is to estimate
the anisotropic stress $\eta$, defined as the ratio of the time-time
and the space-space metric linear potentials $\Phi, \Psi$. A value
of $\eta\not=1$ would signal a modification of gravity (for instance,
a fifth force induced by a scalar field) or the presence of a relativistic
dark matter component. In this paper we have employed a vast collection
of recently available data, from Supernovae Ia to cosmic chronometers,
from lensing to redshift space distortions, to estimate the anisotropic stress
through the statistics $\eta_{\rm{obs}}$,  proposed in \cite{Amendola2012},
that is independent of assumptions about background cosmology, 
 galaxy bias,  initial conditions, and matter abundance.

Since the current datasets have been obtained at different
redshifts, and because $\eta_{\rm{obs}}$ requires derivatives
of data points, we need to interpolate the data in order to build $\eta_{\rm{obs}}$.
We adopted three different strategies to do so: binning, Gaussian
Process, and polynomial regression. 
The Gaussian Process makes the strongest assumption, reducing the uncertainty
to a very small number of parameters. Indeed, the Gaussian Process
method delivers the most stringent error bars. The polynomial 
regression method employs free polynomials in which the order is given by the quality and quantity of data points (up to sixth order, in our case) to fit
the data and evaluate the derivatives at the required points. 

We find that the results are compatible
with each other for the first bin, where data are more abundant. In the second bin, the binning method
is 1.5$\sigma$ away from the GP. For the third and farthest bin, the errors are so large that the comparison
is hardly significant.  In some cases, the standard gravity value $\eta=1$ is two or even three sigma away from our result, but it is in every bin compatible for at least two of the three methods. 

We quote as our fiducial result the error bars produced
by the polynomial regression method: we find
$\eta_{\rm obs}=0.44 \pm 0.92$
at $z=0.294$, $\eta_{\rm obs}=0.42\pm0.89$ at $z=0.58$, and $\eta_{\rm obs}=-0.14\pm3.01$
at $z=0.86$. Assuming a constant $\eta_{\rm obs}$ in this range and
performing a simple weighted average, we find finally $\eta_{\rm obs}=0.49 \pm 0.69$. We consider this
as the most reliable and conservative result.
The other two methods give for a constant anisotropic stress $\eta_{\rm obs}=0.15 \pm 0.27$ (binning) and
$\eta_{\rm obs}=0.62 \pm 0.19$ (Gaussian Process). 

Future surveys, such as the Euclid satellite, will soon produce very large
datasets for all the relevant observables that enter $\eta_{\rm obs}$.
The forecasts produced in \cite{Amendola2013} show that a constant $\eta_{\rm obs}$
could be measured up to a few percent. This is an exciting prospect, if one compares it to
the 100\% (or larger) error bars we find for the present observations. 

\input{rawdatatables.tex}

\section*{Acknowledgments}

A.M.P. gratefully acknowledges the support by the Landesgraduiertenförderung
(LGF) grant of the Graduiertenakademie Universität Heidelberg. 
S.C. acknowledges support from CNRS and CNES grants.
We acknowledge support from DFG through project TR33 ``The Dark Universe''.
 We also acknowledge partial support
from DAAD PPP Portugal bilateral project.
We thank Adriá Gomez for useful discussions on the data analysis.

\appendix

\section{The impact of the $H_0$ choice}

\label{app.H0}

Since our model-independent estimation for $\eta$ requires the dimensionless Hubble function, that is $E(z) = H(z)/H_0$, we need to choose the value of $H_0$ to transform the $H(z)$ data into $E(z)$. However, there is a statistically significant tension between the values measured by different probes, namely the value from the 2018 results of the Planck collaboration \cite{Planck2018} which is 
$H_{0}^{Planck}=67.36\pm0.54\,[\textrm{km/s/Mpc}]$ and the value from HST collaboration \cite{2018ApJ...855..136R} which is 
$H_{0}^{HST}=73.45\pm1.66\,[\textrm{km/s/Mpc}]$. Table \ref{tab.H0comp} describes how the estimation of $\eta$ shifts with the choice of $H_0$. Generally the mean value and uncertainty do not significant change, as previously mentioned. However, for the last bin case, in particular for the binning case, the uncertainty increases for a factor of 10. While this result is compatible with the other bins, it shows how sensitive the binning method is. Since our aim is to have a model-independent estimation of $\eta$, we chose the HST collaboration $H_0$ value as it is approximately independent of a cosmological model.

\input{comparisonH0.tex}

\section{Subtleties of the Gaussian Process method}

\label{app.GP}

We would like to note that the Gaussian Process method and the GaPP code are very sensitive to the dataset that one aims to reconstruct. Using the GaPP code which by default maximizes the hyperparameters, we have performed further tests with the $E(z)$ data, $f \sigma_8$ and $\ln (f \sigma_{8}(z))$. \\

\begin{figure}[htbp]
\centering
\includegraphics[width=0.47\textwidth]{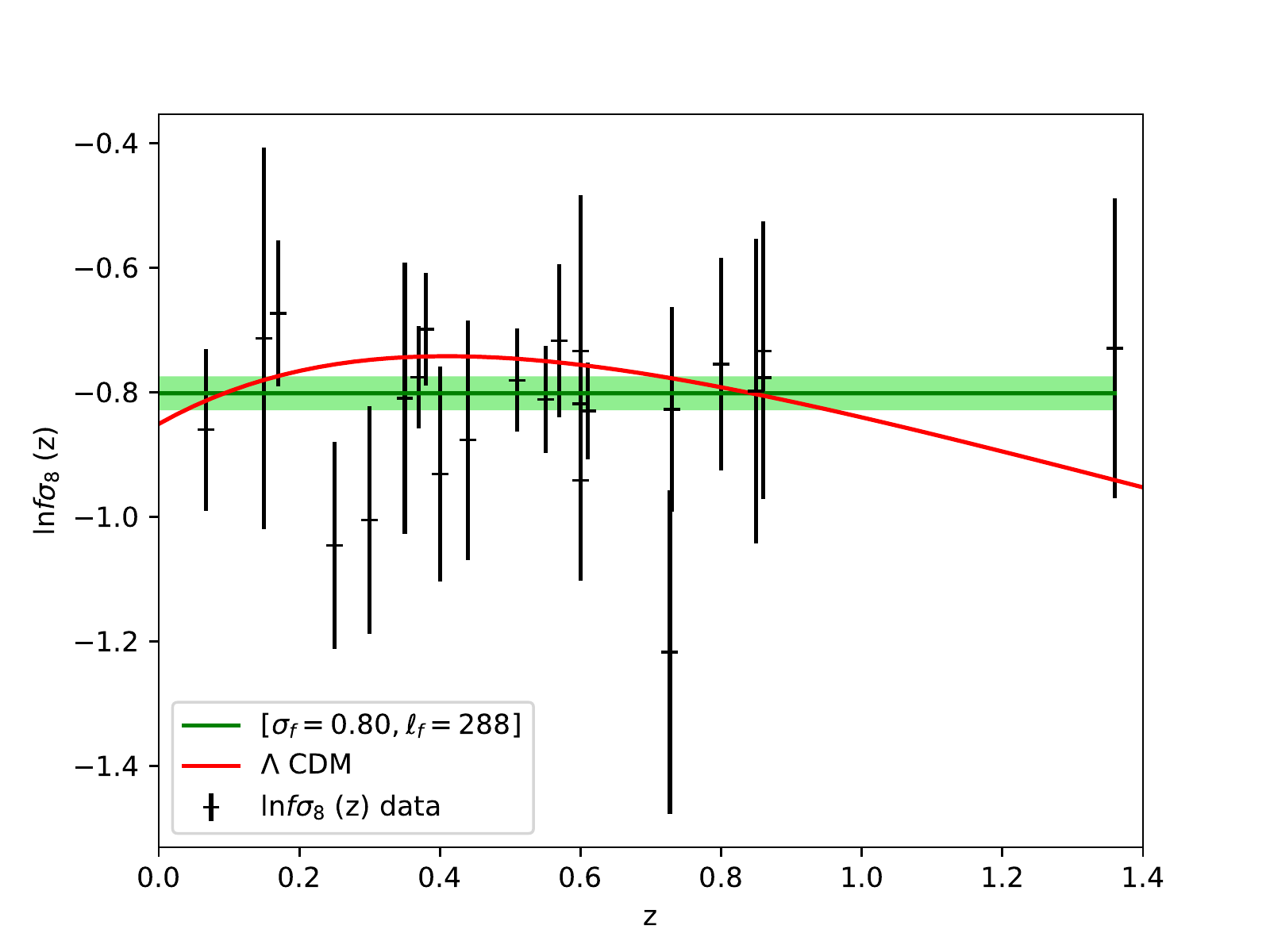}
\includegraphics[width=0.47\textwidth]{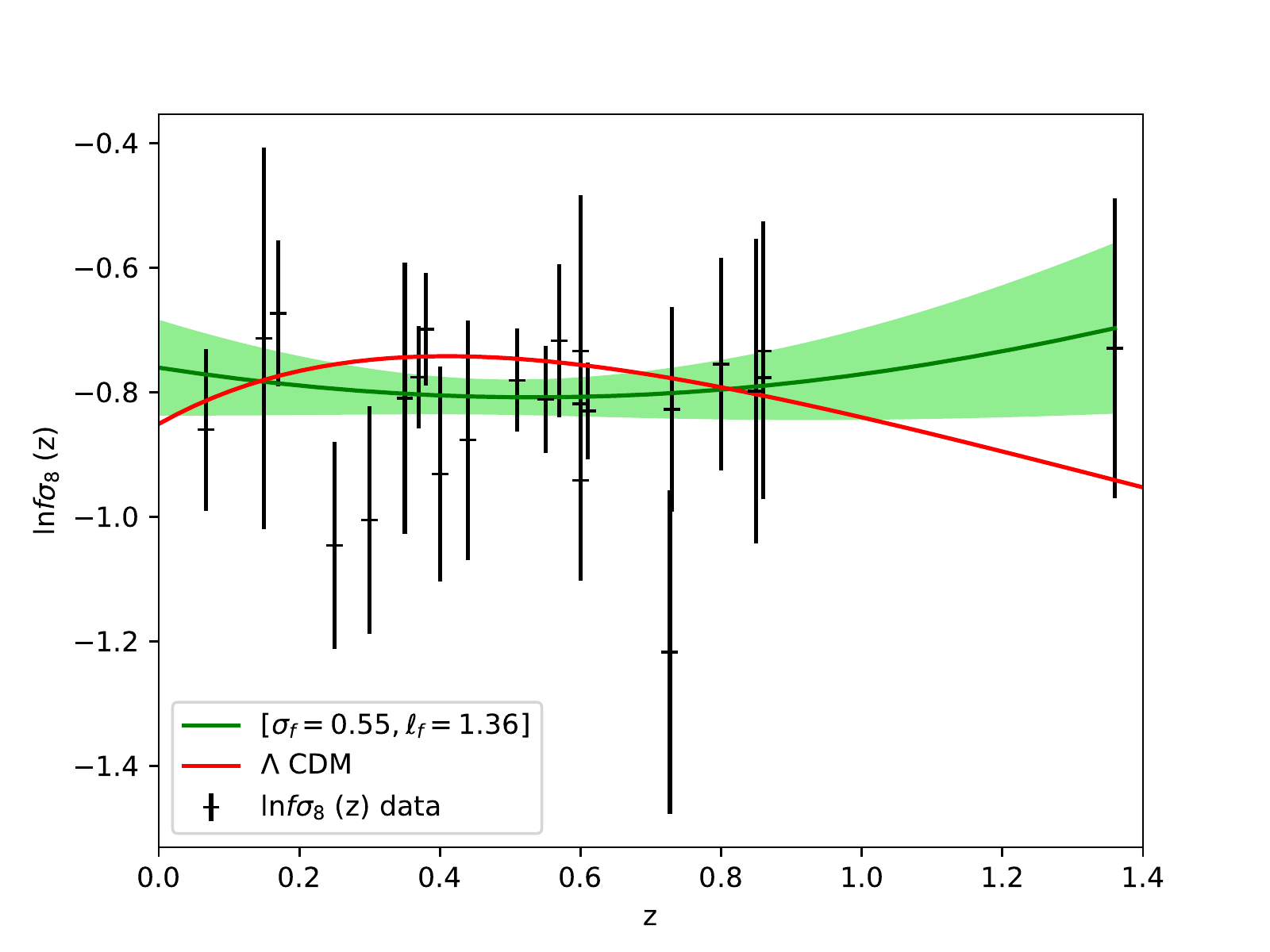}
\caption{Reconstrunction of the $ \ln (f \sigma_{8}(z))$ data by the Gaussian
Process method using the hyperparameters obtained by the GaPP code (left panel)
and using a grid in hyperparameter space with a prior on $\ell_f$ (right panel).}
\label{plt.GaPPtests} 
\end{figure}

In Section \ref{resupart}, we explained another approach to the usage of this code, where instead of letting the code determine the best-fit hyperparameters, we computed the logarithm of the marginal likelihood on a grid of hyperparameters and then we found the values that maximized it. This grid has 300 linear spaced values from 0.01 to 2 for $\sigma_f$ and from 0.01 to the maximum reshift of the dataset for $\ell_f$.
For the $E(z)$ data, we find that there is no significant change but for the $ f \sigma_{8}(z)$ and $ \ln (f \sigma_{8}(z))$ data different reconstructions arise. 
For our work we used the $ \ln(f \sigma_{8}(z))$ data, therefore we show in figure \ref{plt.GaPPtests} the reconstructed function an its uncertainty band in green, together with the data used. On the left side, we see that GaPP estimates a best-fit correlation length of $\ell_f=288$, which yields a very flat reconstruction of  $ \ln (f \sigma_{8}(z)) $. 
On the right side of figure \ref{plt.GaPPtests}, we see that by setting a prior on the correlation length, that is from 0.01 to 1.36, we recover a function that follows much better the general data trend with $\ell_f=1.36$ then the hyperparameters set by the GaPP optimization routine.
\\

\section{Details of the Polynomial Regression Method for the reconstruction of $\eta_{{\rm obs}}$}

\label{app.LRG}

We want to estimate the value of the functions $y^{(i)},$ for $i=1,2,3,4$
at a number of arbitrary points, labeled by subscripts $A,B,C\dots$,
which we can call the interpolated points and assume that there is
a domain $\mathcal{D}$ common to all datasets, in which all the interpolated
points are contained. We further assume that the three initial datasets,
$y^{(j)}$, for $j=0,1,2$ are independent of each other. Now we use
for all initial datasets, a polynomial of the form 
\begin{equation}
g^{(j)}=\sum_{\alpha=0}^{N_{j}}(1+x)^{\alpha}
\end{equation}
with $N_{j}$ the maximum order of the polynomial, which depends on
the characteristics of each dataset $y^{(j)}$, and will be explained
further below. Then the function $f$ in Eq.~(\ref{eq:linearfunc})
will have $N_{j}+1$ coefficients, ranging from $\bar{A}_{0}$ to
$\bar{A}_{N_{j}}$. If we now take the derivative of this function,
we obtain 
\begin{equation}
f^{(j)'}=\sum_{\mu=1}^{N_{j}}-\mu\bar{A}_{\mu}^{(j)}g_{\mu}^{(j)}=-\sum_{\mu=1}^{N_{j}}\bar{B}_{\mu}^{(j)}g_{\mu}^{(j)}
\end{equation}
where $g_{\alpha}^{(j)}$ is the $\alpha$-th term in the sum $g^{(j)}$.
For notational simplicity we define the indices $\alpha,\,\beta$
to always run from 0 to $N_{j}$, while the indices $\mu,\,\nu$ will
run from 1 to $N_{j}$. As we can see, the derivative functions $f^{(j)'}$
have one coefficient less, because there is no $A_{0}$ coefficient.
The relation between the old and new coefficients is 
\begin{equation}
\bar{B}_{\mu}=\mu\bar{A}_{\mu}
\end{equation}
This means that the covariance matrix $(F^{j})^{-1}$ of the coefficients
$\bar{A}^{j}$ has to be modified with a Jacobian of the form 
\begin{equation}
J_{\mu\alpha}^{j}=\frac{\partial\bar{B}_{\mu}}{\partial\bar{A}_{\alpha}}=\mu\delta_{\alpha\mu}=\text{diag}(0,1,2,...,\,N_{j})
\end{equation}
to obtain the covariance matrix $\tilde{F}$ of the new coefficients
\begin{equation}
(\tilde{F}^{j})_{\mu\nu}^{-1}=J_{\mu\alpha}^{j}(F_{\alpha\beta}^{j})^{-1}J_{\beta\nu}\quad.
\end{equation}
Since $\alpha,\beta=0,\dots,N_{j}$ and $\mu,\nu=1,\dots,N_{j}$ the
Jacobian is a rectangular matrix of dimensions $(N_{j}-1)\times N_{j}$,
therefore the $\tilde{F}$ matrices will have a dimension equal to
the original $F$ minus unity.

Summarizing, we will have the following four functions at the wanted
points $A$ 
\begin{equation}
\bar{f}_{A}^{(a)}=\bar{B}_{\{\alpha,\mu\}}^{(a)}p_{A\{\alpha,\mu\}}^{(a)}\label{eq:bestfitf}
\end{equation}
Where due to the derivative, we will have the following basis
functions, 
\begin{align}
p_{\alpha}^{(1)} & =g_{\alpha}^{(1)} & \quad p_{\alpha}^{(2)} & =g_{\alpha}^{(2)} & \quad p_{\mu}^{(3)} & =-g_{\mu}^{(0)} & \quad p_{\mu}^{(4)} & =-g_{\mu}^{(1)}
\end{align}
for $\alpha=0,\dots,N_{j}$ and $\mu=1,\dots,N_{j}$. Which in turn
leads to a change in the vector of coefficients, such that they read
now 
\begin{align}
\bar{B}_{\alpha}^{(1)} & =\bar{A}_{\alpha}^{(1)} & \quad\bar{B}_{\alpha}^{(2)} & =\bar{A}_{\alpha}^{(2)} & \quad\bar{B}_{\mu}^{(3)} & =\mu\bar{A}_{\mu}^{(0)} & \quad\bar{B}_{\mu}^{(4)} & =\mu\bar{A}_{\mu}^{(1)}\quad.
\end{align}

The Fisher matrices for $\bar{B}^{(1)}$ and $\bar{B}^{(2)}$ , are
$F^{(1)}$ and $F^{(2)}$, respectively. For $\bar{B}^{(3)}$ the
Fisher matrix is $\tilde{F}^{(3)}$, while for $\bar{B}^{(4)}$ it
is $\tilde{F}^{(4)}$. The $\tilde{F}$ matrices have a dimension
smaller by one unit than the original $F$.

\begin{align}
C_{\alpha\beta}^{(1,1)} & =\mathrm{Var}(\bar{B}_{\alpha}^{(1)}\bar{B}_{\beta}^{(1)})=\left(F^{(1)}\right)_{\alpha\beta}^{-1}\\
C_{\alpha\beta}^{(2,2)} & =\mathrm{Var}(\bar{B}_{\alpha}^{(2)}\bar{B}_{\beta}^{(2)})=\left(F^{(2)}\right)_{\alpha\beta}^{-1}\\
C_{\mu\nu}^{(3,3)} & =\mathrm{Var}(\bar{B}_{\mu}^{(3)}\bar{B}_{\nu}^{(3)})=\left(\tilde{F}^{(3)}\right)_{\mu\nu}^{-1}\\
C_{\mu\nu}^{(4,4)} & =\mathrm{Var}(\bar{B}_{\mu}^{(4)}\bar{B}_{\nu}^{(4)})=\left(\tilde{F}^{(4)}\right)_{\mu\nu}^{-1}\\
C_{\alpha\beta}^{(1,4)} & =\mathrm{Var}(\bar{B}_{\alpha}^{(1)}\bar{B}_{\beta}^{(4)})=\mathrm{Var}(\bar{A}_{\alpha}^{(1)}\beta\bar{A}_{\beta}^{(1)})=\left(F^{(1)}\right)_{\alpha\gamma}^{-1}J_{\gamma\beta}
\end{align}
The full matrix $\mathcal{C}_{ab,AB}$ is our final result: the covariance
matrix at any two different points $x_{A},\,\,x_{B}$ for any pairs
of datasets $f^{(a)},f^{(b)}$

\begin{equation}
\mathcal{C}_{ab,AB}=C_{\alpha\beta}^{(a,b)}p_{A\alpha}^{(a)}p_{B\beta}^{(b)}
\end{equation}

\bibliographystyle{JHEP}
\bibliography{scaling_bib,refs}

\end{document}

%% file: allfinaldata.tex
\begin{table}[h]
	\centering
	\scalebox{0.75}{
	\begin{tabular}{c|c|ccc|c}
		\hline
		Method  	&	Parameter		& 							& Redshift bins					& 					& Weighted mean \\
		\qquad		&					& $z_1 = 0.294$ 			& $z_2 = 0.58$ 				& $z_3 = 0.86$ 			& \\ 
		\hline \hline		
					& $E(z)$ 			& $1.12 \pm 0.01$ 			& $1.27 \pm 0.02$ 			& $1.51 \pm 0.02$ 		& \\
					& $E'(z)$ 			& $-0.56 \pm 0.07$ 			& $-0.60 \pm 0.36$ 			& $-1.75 \pm 0.66$ 		& \\
		Binning		& $P_2(z)$ 			& $0.75 \pm 0.10$ 			& $0.54 \pm 0.07$ 			& $0.18 \pm 0.14$ 		& \\
					& $P_3(z)$	 		& $-0.17 \pm 0.35$ 			& $0.53 \pm 0.61$ 			& $-1.27 \pm 1.52$ 		& \\
					& $\eta_{obs}(z)$ 	& $0.48 \pm 0.45$ 			& $-0.03 \pm 0.34$			& $-2.78 \pm 6.84$ 		& $0.15 \pm 0.27$ \\
		\hline
					& $E(z)$ 			& $1.10 \pm 0.01 $ 			& $1.30 \pm 0.02 $ 			& $1.55 \pm 0.03 $ 		& \\
					& $E'(z)$ 			& $-0.73 \pm 0.05 $ 		& $-1.30 \pm 0.10 $ 		& $-1.89 \pm 0.16 $ 	& \\
		Gaussian Process& $P_2(z)$ 		& $ 0.74 \pm 0.09 $ 		& $ 0.53 \pm 0.06 $ 		& $ 0.23 \pm 0.11 $ 	& \\
					& $P_3(z)$	 		& $ -0.10 \pm 0.20 $ 		& $ -0.03 \pm 0.21 $ 		& $ -0.21 \pm 0.30 $ 	& \\
					& $\eta_{obs}(z)$ 	& $ 0.38 \pm 0.23 $ 		& $ 0.91 \pm 0.36 $			& $ 0.58 \pm 0.93 $ 	& $0.53 \pm 0.19$ \\
		\hline
							& $E(z)$ 			& $1.12 \pm 0.01$ 		& $1.29 \pm 0.02 $ 		& $1.50 \pm 0.02 $ 		& \\
							& $E'(z)$ 			& $-0.73 \pm 0.04$ 		& $-1.06 \pm 0.04 $ 	& $-1.45 \pm 0.04 $ 	& \\
		Polynomial Regression	& $P_2(z)$ 		& $0.76 \pm 0.15$ 		& $ 0.55 \pm 0.15 $ 	& $ 0.18 \pm 0.14 $ 	& \\
							& $P_3(z)$			& $-0.09 \pm 0.80$ 		& $ 0.14 \pm 0.78 $		& $ -0.17 \pm 3.02 $ 	& \\
							& $\eta_{obs}(z)$ 	& $0.57 \pm 1.05$ 		& $ 0.48 \pm 0.96 $ 	& $-0.11 \pm 3.21 $ 	& $0.49 \pm 0.69$\\
		\hline
	\end{tabular}}
	\caption{The reconstructed or measured model-independent variables $E, E', P_2, P_3, \eta(z)$ at three different redshifts $z=(0.294, 0.58, 0.86)$, together with their $1\sigma$ errors, for each of the reconstruction methods. The polynomial regression method is compabitle with the $\Lambda$CDM scenario while the other two methods show some tension at lower redshift.}
	\label{tab.results} 
\end{table}

%% file: rawdatatables.tex
\begin{table}[!htbp] 
	\centering %
	\caption{$H(z)$ measurements compiled by \cite{Yu2017} with the respective original references.}	
	\scalebox{0.75}{
	\begin{tabular}{c|c|c|c|c||}
		\hline 
		$z$  	& $H(z)$ 			& $\sigma_{H(z)}$ & Reference & Method \tabularnewline 
			& (km/s/Mpc) 	& (km/s/Mpc) 	&	& \tabularnewline
		\hline 	\hline 
		0.07  	& 69  	& 19.6  & \cite{Zhang2012} 		& 1 \tabularnewline
		0.09  	& 69  	& 12  	& \cite{Simon2004} 		& 1 \tabularnewline
		0.12  	& 68.6  & 26.2  & \cite{Zhang2012} 	& 1 \tabularnewline
		0.17  	& 83  	& 8  	& \cite{Simon2004} 		& 1 \tabularnewline
		0.179  	& 75  	& 4  	& \cite{Moresco:2012jh} & 1 \tabularnewline
		0.199  	& 75  	& 5  	& \cite{Moresco:2012jh} & 1 \tabularnewline
		0.2  	& 72.9  & 29.6  & \cite{Zhang2012}		& 1 \tabularnewline
		0.27  	& 77  	& 14  	& \cite{Simon2004} 		& 1 \tabularnewline
		0.28  	& 88.8  & 36.6  & \cite{Zhang2012} 	& 1 \tabularnewline
		0.352  	& 83  	& 14  	& \cite{Moresco:2012jh} & 1 \tabularnewline
		0.38  	& 81.5  & 1.9  	& \cite{Alam2016} 		& 2 \tabularnewline
		0.3802  & 83  	& 13.5  & \cite{Moresco2016} 	& 1 \tabularnewline
		0.4  	& 95  	& 17  	& \cite{Simon2004} 		& 1 \tabularnewline
		0.4004  & 77  	& 10.2  & \cite{Moresco2016} 	& 1 \tabularnewline
		0.4247  & 87.1  & 11.2 & \cite{Moresco2016} 	& 1 \tabularnewline
		0.44  	& 82.6  & 7.8  & \cite{Blake2012} 		& 2 \tabularnewline
		0.4497  & 92.8  & 12.9 & \cite{Moresco2016} 	& 1 \tabularnewline
		0.4783  & 80.9  & 9  	& \cite{Moresco2016} 	& 1 \tabularnewline
		0.480  	& 97  	& 62  	& \cite{Stern2009} 		& 1 \tabularnewline
		\hline
		\end{tabular}
		\begin{tabular}{c|c|c|c|c}
		\hline 
		$z$  	& $H(z)$ 			& $\sigma_{H(z)}$ & Reference & Method \tabularnewline 
			& (km/s/Mpc) 	& (km/s/Mpc) 	&	& \tabularnewline
		\hline 	\hline
		0.510  	& 90.4  & 1.9  	& \cite{Alam2016} 		& 2 \tabularnewline
		0.593  	& 104  	& 13  	& \cite{Moresco:2012jh} & 1 \tabularnewline
		0.600  	& 87.9  & 6.1  	& \cite{Blake2012} 		& 2 \tabularnewline
		0.610  	& 97.3  & 2.1  	& \cite{Alam2016} 		& 1 \tabularnewline
		0.680  	& 92  	& 8  	& \cite{Moresco:2012jh}	& 1 \tabularnewline
		0.730  	& 97.3  & 7  	& \cite{Blake2012} 		& 2 \tabularnewline
		0.781  	& 105  	& 12  	& \cite{Moresco:2012jh} & 1 \tabularnewline
		0.875  	& 125  	& 17  	& \cite{Moresco:2012jh} & 1 \tabularnewline
		0.880  	& 90  	& 40  	& \cite{Stern2009} 		& 1 \tabularnewline
		0.900  	& 117  	& 23  	& \cite{Simon2004} 		& 1 \tabularnewline
		1.037  	& 154  	& 20  	& \cite{Moresco:2012jh} & 1 \tabularnewline
		1.300  	& 168  	& 17  	& \cite{Simon2004} 		& 1 \tabularnewline
		1.363  	& 160  	& 33.6  & \cite{Moresco:2015cya} & 1 \tabularnewline
		1.430  	& 177  	& 18  	& \cite{Simon2004} 		& 1 \tabularnewline
		1.530  	& 140  	& 14  	& \cite{Simon2004} 		& 1 \tabularnewline
		1.750  	& 202  	& 40  	& \cite{Simon2004} 		& 1 \tabularnewline
		1.965  	& 186.5 & 50.4  & \cite{Moresco:2015cya} 	& 1 \tabularnewline
		2.340  	& 222  	& 7  	& \cite{Delubac:2014aqe} 	& 3 \tabularnewline
		2.360  	& 226  	& 8  	& \cite{Font-Ribera:2013wce} & 3 \tabularnewline
		\hline 
	\end{tabular}}
	\label{tab.hzdata} 
\end{table}

\begin{table}[!htbp]
	\centering
	\caption{$E(z)$ measurements from \cite{Riess2017}. The error of the last measurement is not symmetric therefore it
		was recalculated as the quadrature of the $1 \sigma$ bounds on the left and right side of the central value.}
	\begin{tabular}{c|c|c}
		\hline 
		$z$  & $E(z)$  & $\sigma_{E(z)}$ \tabularnewline
		\hline 
		\hline 
		0.07  & 0.997  & 0.023 \tabularnewline
		0.2  & 1.111  & 0.020 \tabularnewline
		0.35  & 1.128  & 0.037 \tabularnewline
		0.55  & 1.364  & 0.063 \tabularnewline
		0.9  & 1.52  & 0.12 \tabularnewline
		1.5  & 2.67  & 0.675 \tabularnewline
		\hline 
	\end{tabular}
\label{tab.ezdata} 
\end{table}

\begin{table}[!htbp]
	\scalebox{0.9}{
    \begin{minipage}{.5\linewidth}
      \caption{Covariance matrix for the $H(z)$ data from \cite{Alam2016}.}
      \centering
	\begin{tabular}{c|ccc}
		\hline
		z 		& & Covariance matrix & \\
		\hline \hline
		0.38 	& 3.65 & 1.78 & 0.93\\
		0.51 	& 1.78 & 3.65 & 2.20\\
		0.61	& 0.93 & 2.20 & 4.45\\
		\hline
	\end{tabular}
	\label{tab.covAlam}
    \end{minipage}%
    \begin{minipage}{.5\linewidth}
      \centering
        \caption{Covariance matrix for the $H(z)$ data from \cite{Blake2012}.}
		\begin{tabular}{c|ccc}
			\hline
			z 		& 		 	& Covariance matrix & \\
			\hline \hline
			0.44 	& 0.0064 	& 0.0025704		& 0 \\
			0.60 	& 0.0025704	& 0.003969		& 0.00254016 \\
			0.73	& 0			& 0.00254016	& 0.005184 \\
			\hline
		\end{tabular}
		\label{tab.covWig}
    \end{minipage}
    }
\end{table}

\begin{table}[!htbp] 
	\centering %
	\caption{Table of the $E_{g}(z)$ data set. The first column is the redshift, the second one is the value with the corresponding error on the third column. The forth column shows the considered interval in real space that was used to obtain each data point and the last column points to the reference in the literature.}
	\begin{tabular}{c|c|c|c|c}
		\hline 
		z  & $E_{g}(z)$  & $\sigma_{E_{g}(z)}$  & Scale $(h^{-1} Mpc)$& Reference \tabularnewline
		\hline 
		\hline
		0.267 & 0.43  & 0.13  & $5 < R_p < 40$ & \cite{Amon2017} \\
		0.305 &  0.27 &   0.08 & $ 5 < R_p < 60 $ & \cite{Amon2017} \\
		0.32  & 0.40  & 0.09  & $R_p > 3 $ & \cite{Blake2016} \tabularnewline
		0.32  & 0.48  & 0.10  & $R_p > 10 $ & \cite{Blake2016} \tabularnewline
		0.554 &  0.26 &   0.07 & $ 5 < R_p < 60 $ & \cite{Amon2017} \\
		0.57  & 0.31  & 0.06  & $R_p > 3 $ & \cite{Blake2016} \tabularnewline
		0.57  & 0.30  & 0.07  & $R_p > 10 $ & \cite{Blake2016} \tabularnewline
		0.60  & 0.16  & 0.09  & $3 < R_p < 20$ & \cite{DelaTorre2016} \tabularnewline
		0.86  & 0.09  & 0.07  & $3 < R_p < 20$ & \cite{DelaTorre2016} \tabularnewline 
		\hline 
	\end{tabular}
	\label{tab.egdata} 
\end{table}

\begin{table}[!htbp] 
	\centering %
	\caption{$f\sigma_{8}(z)$ data with the correspondent redshift and error. The fourth column points to the reference in literature.}
	\scalebox{0.8}{
	\begin{tabular}{c|c|c|c||}
		\hline 
		z  & $f\sigma_{8}(z)$  & $\sigma_{f\sigma_{8}(z)}$  & Reference \tabularnewline
		\hline 
		\hline
		0.067  	& 0.423  	& 0.055  & \cite{Beutler2012} \tabularnewline		
		0.15  	& 0.49  	& 0.15  & \cite{Howlett2015} \tabularnewline
		0.17 	& 0.51 		& 0.06 	& \cite{Song2008} \tabularnewline
		0.25  	& 0.3512  	& 0.0583  & \cite{Samushia2012} \tabularnewline
		0.30 	& 0.366 	& 0.067 & \cite{Tojeiro:2012rp} \tabularnewline
		0.35   	& 0.445		& 0.097  & \cite{Chuang2012} \tabularnewline
		0.37  	& 0.4602  	& 0.0378  & \cite{Samushia2012} \tabularnewline
		0.38 	& 0.497 	& 0.045 & \cite{Alam2016} \tabularnewline		
		0.40	& 0.394		& 0.068	& \cite{Gil-Marin:2015sqa} \tabularnewline	
		\hline
		\end{tabular}
		\begin{tabular}{c|c|c|c||}
		\hline 
		z  & $f\sigma_{8}(z)$  & $\sigma_{f\sigma_{8}(z)}$  & Reference \tabularnewline
		\hline 
		\hline
		0.44  	& 0.416  	& 0.080  & \cite{Blake2012} \tabularnewline
		0.51 	& 0.458 	& 0.038 & \cite{Alam2016} \tabularnewline
		0.55	& 0.444     & 0.038 & \cite{Gil-Marin:2015sqa} \tabularnewline
		0.57  	& 0.488  	& 0.060  & \cite{Gil-Marin2016}, \cite{Chuang2016} \tabularnewline
		0.60  	& 0.390  	& 0.063  & \cite{Blake2012} \tabularnewline		
		0.60 	& 0.441 	& 0.071 & \cite{Tojeiro:2012rp} \tabularnewline		
		0.60  	& 0.48  	& 0.11  & \cite{delaTorre:2016rxm} \tabularnewline		
		0.60  	& 0.48  	& 0.12  & \cite{delaTorre:2016rxm} \tabularnewline
		0.61 	& 0.436 	& 0.034 & \cite{Alam2016} \tabularnewline	
		\hline 
		\end{tabular}
		\begin{tabular}{c|c|c|c}
		\hline 
		z  & $f\sigma_{8}(z)$  & $\sigma_{f\sigma_{8}(z)}$  & Reference \tabularnewline
		\hline 
		\hline
		0.727  	& 0.296  	& 0.077  & \cite{Hawken2016} \tabularnewline		
		0.73  	& 0.437  	& 0.072  & \cite{Blake2012} \tabularnewline
		0.80  	& 0.47  	& 0.08  & \cite{DelaTorre2013} \tabularnewline
		0.85  	& 0.45  	& 0.11  & \cite{Mohammad2017} \tabularnewline
		0.86 	& 0.46  	& 0.09  & \cite{delaTorre:2016rxm} \tabularnewline
		0.86 	& 0.48  	& 0.10  & \cite{delaTorre:2016rxm} \tabularnewline
		1.36  	& 0.482  	& 0.116  & \cite{Okumura2015} \tabularnewline
				& 			& 		& \tabularnewline
				& 			&		& \tabularnewline
		\hline 
		\end{tabular}
		}
	\label{tab.fs8} 
\end{table}

%% file: comparisonH0.tex
\begin{table}[h]
	\centering
	\begin{tabular}{c|c|ccc}
		Method  			& Choice of $H_0$ 	& 					& $\eta(z)$			& 					\\
		\qquad				&					& $z_1 = 0.294$ 	& $z_2 = 0.58$ 			& $z_3 = 0.86$ 		\\
		\hline
		Binning 			& $H_0$ HST			& $0.48 \pm 0.45$ 	& $-0.03 \pm 0.34$		& $-2.78 \pm 6.84$ 	\\
		\qquad				& $H_0$ Planck 2018	& $0.56 \pm 0.54$	& $-0.14 \pm 0.32$		& $-6.75 \pm 75.64$ \\
		\hline
		GaPP 				& $H_0$ HST			& $0.49 \pm 0.25$ 	& $0.94 \pm 0.33$		& $0.27 \pm 0.67$ \\
		\qquad				& $H_0$ Planck 2018	& $0.31 \pm 0.22$	& $0.72 \pm 0.33$		& $0.36 \pm 0.79$ \\
		\hline		
		Linear Regression 	& $H_0$ HST			& $0.57 \pm 1.05$ 	& $0.48 \pm 0.96 $ 		& $-0.11 \pm 3.21 $ \\
		\qquad				& $H_0$ Planck 2018	& $0.51 \pm 1.07$ 	& $0.37 \pm 0.93 $ 		& $-0.18 \pm 3.11 $ \\
		\hline \hline
	\end{tabular}
	\caption{The reconstructed $\eta(z)$ using different values of $H_0$ to normalize the $H(z)$ data at three different redshifts $z=(0.294, 0.58, 0.86)$ with its respective $1\sigma$ errors, for each of the reconstruction methods.}
	\label{tab.H0comp} 
\end{table}